\documentclass[preprintnumbers, prd, showpacs, floatfix,onecolumn, superscriptaddress,nofootinbib]{revtex4}
\usepackage{amsfonts}
\usepackage{amsmath}
\usepackage{amssymb,epsf}
\usepackage{latexsym}
\usepackage{graphicx,epsfig}
\usepackage{amssymb}
\usepackage{subfigure}
\usepackage{epstopdf}

\begin{document}

\title{Counterterm method in dilaton gravity and the critical behavior of
dilaton black holes with power-Maxwell field}
\author{Z. Dayyani$^{1}$, A. Sheykhi $^{1,2}$\footnote{
asheykhi@shirazu.ac.ir} and M. H. Dehghani$^{1,2}$\footnote{
mhd@shirazu.ac.ir}}
\address{$^1$ Physics Department and Biruni Observatory, College of
Sciences, Shiraz University, Shiraz 71454, Iran\\
         $^2$Research Institute for Astronomy and Astrophysics of Maragha
         (RIAAM), P.O. Box 55134-441, Maragha, Iran}
\begin{abstract}
We investigate the critical behavior of an $(n+1)$-dimensional topological
dilaton black holes, in an extended phase space in both canonical and
grand-canonical ensembles, when the gauge field is in the form of
power-Maxwell field. In order to do this we introduce for the first time the
counterterms that remove the divergences of the action in dilaton gravity
for the solutions with curved boundary. Using the counterterm method, we
calculate the conserved quantities and the action and therefore Gibbs free
energy in both the canonical and grand-canonical ensembles. We treat the
cosmological constant as a thermodynamic pressure, and its conjugate
quantity as a thermodynamic volume. In the presence of power-Maxwell field,
we find an analogy between the topological dilaton black holes with van der
Walls liquid-gas system in all dimensions provided the dilaton coupling
constant $\alpha$ and the power parameter $p$ are chosen properly.
Interestingly enough, we observe that the power-Maxwell dilaton black holes
admit the phase transition in both canonical and grand-canonical ensembles.
This is in contrast to RN-AdS, Einstein-Maxwell-dilaton and
Born-Infeld-dilaton black holes, which only admit the phase transition in
the canonical ensemble. Besides, we calculate the critical quantities and
show that they depend on $\alpha$, $n$ and $p$. Finally, we obtain the
critical exponents in two ensembles and show that they are independent of
the model parameters and have the same values as mean field theory.
\end{abstract}

\maketitle

\section{Introduction}

It has been shown that one can extend the thermodynamic phase space of a
Reissner-Nordstrom (RN) black holes in an anti-de Sitter (AdS) space, by
considering the cosmological constant as a thermodynamic pressure, $%
P=-\Lambda /8\pi $ and its conjugate quantity as a thermodynamic volume \cite%
{Do1,Ka,Do2,Do3,Ce1,Ur}. The studies on the critical behavior of black hole
spacetimes, in a wide range of gravity theories, have got a lot of
enthusiasm. Let us review some works in this direction. For example, $P$-$V$
criticality of charged AdS black holes has been investigated in \cite{MannRN}
and it was shown that indeed there is a complete analogy for RN-AdS black
holes with the van der Walls liquid-gas system. In particular, it was found
that the critical exponents of this system coincide with those of the van
der Waals system \cite{MannRN}. When the gauge field is the Born-Infeld
nonlinear electrodynamics, extended phase space thermodynamics of
charged-AdS black holes have been investigated in \cite{MannBI}. In this
case, one needs to introduce a new thermodynamic quantity conjugate to the
Born-Infeld parameter which is required for consistency of both the first
law of thermodynamics and the corresponding Smarr relation \cite{MannBI}.
The studies were also extended to the rotating black holes. In this regards,
phase transition, critical behavior, and critical exponents of Myers-Perry
black holes have been explored in \cite{Sherkat}. Besides, it was shown that
charged and rotating black holes in three spacetime dimensions do not
exhibit critical phenomena \cite{MannBI}. Other studies on the critical
behavior of black hole spacetimes in an extended phase space have been
carried out in \cite{Sherkat1,Rabin,Zou,Hendi,John,De,Xi}.

It is also of great interest to generalize the study to dilaton gravity \cite%
{Ren}. Critical behavior of black holes in Einstein-Maxwell-dilaton (EMd)
gravity in the presence of Liouville-type potentials, which is regarded as
the generalization of the cosmological constant, has been explored in \cite%
{Kamrani}. Although, the asymptotic behavior of these solutions \cite%
{Kamrani} are neither flat nor AdS, it was observed that the critical
exponents have the universal mean field values and do not depend on the
details of the system, although the thermodynamic quantities depend on the
dilaton coupling constant, $\alpha$ \cite{Kamrani}. The studies were also
extended to the nonlinear electrodynamics in dilaton gravity. In the context
of Einstein-Born-Infeld-dilaton (EBId) gravity, critical behavior of $(n+1)$%
-dimensional topological black holes in an extended phase space was explored
in \cite{dayyani}. By interpreting the constant $\Lambda $ and the BI
parameter $\beta $ as thermodynamic quantities, it was shown that the the
phase space can be enlarged. It was also argued that although thermodynamic
quantities depend on the dilaton coupling constant, BI parameter and the
dimension of the spacetime, they are universal and are independent of metric
parameters \cite{dayyani}.

One may also interested in studying the critical behavior of the dilaton
black holes when the gauge field is in the form of the power-Maxwell field.
There are some motivations for studying the critical behavior of the
nonlinear power-Maxwell Lagrangian, instead of the usual linear Maxwell
case. The first reason comes from the fact that, while the Maxwell
Lagrangian is only conformally invariant in four dimensions, the
power-Maxwell field is conformally invariant in $(n+1)$-dimensional
spacetime for $p=(n+1)/4$, where $p$ is the power parameter of the
Power-Maxwell Lagrangian. Besides, it is worthwhile to investigate the
effects of exponent $p$ on the critical behavior of the black holes and see
whether it can change the $P$-$V$ criticality and the critical exponents of
the system or not. The investigations on the black object solutions coupled
to a conformally invariant Maxwell field have got a lot of attention from
different perspective \cite%
{kord50,kord51,kord52,kord53,kord54,kord55,kord56,kord57,kord58,kord59,kord60}%
. Thermodynamics and thermal stability, in canonical and grand-canonical
ensembles, of higher dimensional topological dilation black holes with
power-Maxwell field have been studied in \cite{Kord}. It was shown that the
solutions exist provided one assumes three Liouville-type potentials for the
dilaton field, and in the case of the Maxwell field, one of the Liouville
potentials vanishes \cite{Kord}.

In this paper, we intend to study the extended thermodynamic phase space in
dilaton gravity by investigating the critical behavior of the $(n+1)$%
-dimensional dilaton black holes coupled to nonlinear power-Maxwell field.
We shall consider both canonical and grand-canonical ensembles. In order to
do this we should calculate the finite action in these ensembles. It is
known that the total action of all the theories of gravity is divergences on
the solutions \cite{BY,23mehdizade,24mehdizade}. Since the action diverges,
all the other conserved quantities which is calculated through the use of
this action are divergent too. One method of removing these divergences is
through the use of the background subtraction method of Brown and York \cite%
{BY}. Such a procedure causes the resulting physical quantities to depend on
the choice of reference background. Especially, in the case of dilaton
gravity, our calculations show that the action calculated by use of the
subtraction method is not correct. Another way of removing the divergences
in the action and conserved quantities is through the use of counterterm
method. In this method one may remove the divergences in the action for
(A)dS solutions by adding counterterms which are functional of the boundary
curvature invariants \cite{25mehdizade,26mehdizade,27mehdizade}. Indeed, by
using this method one can calculate the action and conserved quantities
intrinsically without reliance on any reference spacetime \cite%
{28mehdizade,30mehdizade,31mehdizade}. Due to this fact, the counterterm
method has been applied to many cases such as black holes with rotation, NUT
charge and various topologies \cite{32mehdizade,33mehdizade,34mehdizade}.
Here we want to generalize the counterterm method to the case of Einstein
gravity in the presence of a dilaton field for the solutions with curved
boundary. In Einstein gravity, although there may exist a large number of
possible boundary curvature invariants, only a finite number of them are
non-vanishing on a boundary at infinity. But in the case of dilaton field
coupled to gravity, the asymptotic behavior of the solutions may be neither
(A)dS nor flat and therefore there are an infinite number of non-vanishing
boundary curvature invariant terms at infinity for the case of curved
horizon. Only for black holes with flat horizon, the curvature of the
boundary vanishes and there exist only one term proportional to square root
of the determinant of the boundary metric \cite{Dilaton}. In the case of
solutions with curved boundary and non-AdS asymptotic, due to the fact that
there exist an infinite number of counterterms, it is difficult to use the
counterterm method in order to calculate the finite action. Due to this
fact, this method has not been used till now. In this paper, for the first
time, we introduce the counterterm method for the calculation of finite
action and physical quantities in dilaton black holes with curved horizon.
By using the counterterm method we calculate the action and Gibbs free
energy in both canonical and grand canonical ensembles to study the phase
transition of the system and compare\ them with van der Waals fluid.

This paper is outlined as follows. In Sec. \ref{Count}, we introduce
counterterms in dilaton gravity to get a finite value for mass and action in
both canonical and grand canonical ensembles. In Sec. \ref{PLM}, we present
basic field equations and a class of $(n+1)$-dimensional topological dilaton
black hole solutions coupled to a nonlinear power-Maxwell field and review
their thermodynamic properties. In Sec. \ref{Ph}, we study the phase
structure of the solutions and present the generalized Smarr relation in the
presence of the dilaton field. In Sec. \ref{Can}, we investigate the analogy
of the obtained dilaton black holes with van der Waals liquid-gas system in
the canonical ensemble by fixing charge at infinity. In this ensemble, we
obtain the equation of state, the critical behavior and critical exponents
and study Gibbs free energy of the solutions. In Sec. \ref{Grand}, we
consider the possibility of the phase transition in the grand-canonical
ensemble by fixing the electric potential at infinity and find that in
contrast to the RN-AdS, EMd and EBId black holes, the phase transition
occurs for $p\neq 1$. Then, we calculate the critical exponents and find
that they match to mean field value (the same as the van der Waals liquid).
Also, we obtain the expression for Gibbs free energy in this ensemble and
study its behavior. The last section is devoted to summary and conclusions.

\section{Counterterm method in dilaton gravity}

\label{Count}

In this section we want to introduce the counterterms for Einstain-dilaton
gravity. The action of $(n+1)$-dimensional $(n\geq 3)$ coupled to the
dilaton field can be written as
\begin{equation}
I_{bulk}=-\frac{1}{16\pi }\int d^{n+1}x\sqrt{-g}\left\{ \mathcal{R}-\frac{4}{%
n-1}(\nabla \Phi )^{2}-V(\Phi )\right\} ,  \label{Act1}
\end{equation}%
where $\mathcal{R}$ is the Ricci scalar, $\Phi $ is the dilaton field and $%
V(\Phi )$ is a potential for $\Phi $.The bulk action of Einstein gravity
which is coupled to the dilaton field (\ref{Act1}) does not have a
well-defined variational principle. In order to make it an action with a
well-defined variational principle, one should add the following
Gibbons-Hawking surface term to the bulk action
\begin{equation*}
I_{GH}=-\frac{1}{8\pi }\int_{\partial \mathcal{M}}d^{n}x\sqrt{-\gamma }K,
\end{equation*}%
where $\gamma _{ij}$ and $K$ denote the induced metric and extrinsic
curvature of the boundary $\partial \mathcal{M}$, respectively. It is clear
that the action $I_{bulk}+I_{GH}$ is not finite. As in the case of
Einstein-Hilbert action, one needs to add counterterms to the action to get
a finite value. Since the counterterms should be reduced to those of
Einstein gravity in the absence of dilaton, these counterterms may be
written from the curvature invariants of the boundary metric. The
coefficients of the boundary terms should be chosen such that the divergences
in the bulk are canceled for all possible boundary topologies permitted by
the equations of motion. For asymptotic AdS solutions, we have only a finite
number of counterterms that do not vanish at infinity. Here, because of
considering a dilaton field coupled to gravity, the asymptotic behaviour of
the solutions is not AdS and therefore we may have an infinite numbers of
counterterms which do not vanish at infinity. First we consider the case in
four dimensions.

\subsection{Counterterms in four dimensions}

Defining $l\equiv \lbrack (\alpha ^{2}-3)/\Lambda ]^{1/2}$, the counterterms
in 4 dimensions may be written as:
\begin{equation}
I_{ct}=\frac{1}{8\pi }\int d^{3}x\sqrt{-\gamma }\left\{ \frac{2}{le^{-\alpha
\Phi }}+\frac{le^{-\alpha \Phi }}{2(1-\alpha ^{4})}\mathcal{R}%
-\sum_{s=2}^{\infty }\frac{A_{s}}{2(\alpha ^{4}-1)^{s}}\left( \frac{%
le^{-\alpha \Phi }}{2}\right) ^{2s-1}\mathcal{R}^{s}\right\} ,  \label{Ict}
\end{equation}%
where%
\begin{eqnarray*}
A_{2} &=&1,\text{ \ \ \ \ \ \ \ \ \ \ \ \ \ }\hspace{0.6cm}A_{3}=2, \\
A_{s} &=&A_{s-2}+2A_{s-1};\text{ \ \ \ \ \ }s\geq 4.
\end{eqnarray*}%
It is worth to mention that in the absence of dilaton $(\alpha =0)$ each
term in the summation is zero and therefore the counterterm reduces to that
of AdS solutions \cite{BY}. But, in the presence of dilaton each
term in the summation of Eq. (\ref{Ict}) may not vanish at infinity. As $%
\alpha $ increases, the number of non-vanishing terms in the series will
increase. Of course as Eqs. (\ref{res1}) and (\ref{res2}) show $\alpha $ is
less than 1 in four dimensions. For example for $\alpha <1/3$, all the terms
in the summation vanish and therefore the counterterms of Einstein gravity
remove the divergences in the action while for $1/3\leq \alpha <3/5$ only
the first does not vanish. Indeed, the number of non-vanishing terms in the
summation for $\alpha <(2N-1)/(2N+1)$ is equal to $N$. Fortunately at
infinity, the summation of all the divergent terms in the total action ($%
I_{bulk}+I_{GH}+I_{ct}$), including the summation term of Eq. (\ref{Ict}),
goes to zero as $s$ goes to infinity. Thus, in order to calculate the finite
action, one only needs to calculate the finite terms in the first two terms
of Eq. (\ref{Ict}) in four dimensions. That is, one need to consider only
the generalization of the counterterms of Einstein gravity in the presence
of dilaton:%
\begin{equation*}
I_{total}=\text{Finite terms of}\left\{ I_{bulk}+I_{GH}+\frac{1}{8\pi }\int
d^{3}x\sqrt{-\gamma }\left( \frac{2}{le^{-\alpha \Phi }}+\frac{le^{-\alpha
\Phi }}{2(1-\alpha ^{4})}\mathcal{R}\right) \right\} .
\end{equation*}

Having the total finite action, one can use the Brown and York definition
\cite{BY} to construct a divergence free stress-energy tensor as
\begin{eqnarray}
T^{ab} &=&\frac{1}{8\pi }\left\{ (K^{ab}-K\gamma ^{ab})+\frac{2}{le^{-\alpha
\Phi }}\gamma ^{ab}-\frac{le^{\alpha \Phi }}{(1-\alpha ^{4})}(\mathcal{R}%
^{ab}-\frac{1}{2}\mathcal{R}\gamma ^{ab})\right.  \notag \\
&&\ +\left. \sum_{s=2}\frac{sA_{s}}{(\alpha ^{4}-1)^{s}}\left( \frac{l}{2}%
e^{-\alpha \Phi }\right) ^{2s-1}\left[ \mathcal{R}^{s-1}(\mathcal{R}^{ab}-%
\frac{1}{2s}\mathcal{R}\gamma ^{ab})-(\nabla ^{a}\nabla ^{b}-g^{ab}\nabla
^{2})\mathcal{R}^{s-1}\right] \right\} .  \label{Stress}
\end{eqnarray}

To compute the conserved charges of the spacetime, we choose a spacelike
surface $\Sigma $ in $\partial \mathcal{M}$ with metric $\sigma _{ij}$, and
write the boundary metric in ADM form:
\begin{equation}
\gamma _{ab}dx^{a}dx^{a}=-\mathcal{N}^{2}dt^{2}+\sigma _{ij}\left( d\varphi
^{i}+\mathcal{V}^{i}dt\right) \left( d\varphi ^{j}+\mathcal{V}^{j}dt\right) ,
\end{equation}%
where the coordinates $\varphi ^{i}$ are the angular variables
parameterizing the hypersurface of constant $r$ around the origin, and $%
\mathcal{N}$ and $\mathcal{V}^{i}$ are the lapse and shift functions
respectively. The conserved charges associated to a Killing vector $\xi ^{a}$
is
\begin{equation}
\mathcal{Q}(\xi )=\int_{\Sigma }d^{2}x\sqrt{\sigma }u^{a}T_{ab}\xi ^{b},
\label{Cons4}
\end{equation}%
where $\sigma $ is the determinant of the metric $\sigma _{ij}$ and $u^{a}$
is the normal to the quasilocal boundary hypersurface $\Sigma $.$\ $For
boundaries with timelike Killing vector ($\xi =\partial _{t}$) one obtains
the conserved mass of the system enclosed by the boundary $\Sigma $.
Again the summation of all the divergent terms in the above equation at
infinity is zero and therefore one should consider only the finite terms in
the first three terms of Eq. (\ref{Str}) in order to calculate the conserved
quantities from Eq. (\ref{Cons4}).

\subsection{Counterterms in ($n+1$) dimensions}

Now, it is easy to generalize the counterterms of previous subsection to the
case of ($n+1$) dimensions. Defining
\begin{equation}
l\equiv \left( \frac{(\alpha ^{2}-n)(n-1)}{2\Lambda }\right) ^{1/2},
\label{Lambda}
\end{equation}%
the counterterm may be written as
\begin{eqnarray}
I_{ct} &=&\frac{1}{8\pi }\int d^{n}x\sqrt{-\gamma }\{\frac{n-1}{le^{-\alpha
\Phi }}+\frac{le^{-\alpha \Phi }}{2(1-\alpha ^{2})(n-2+\alpha ^{2})}\mathcal{%
R}+\frac{l^{3}e^{-3\alpha \Phi }}{2(n-4)(1-\alpha ^{2})(n-2+\alpha ^{2})^{2}}%
[\mathcal{R}_{ab}\mathcal{R}^{ab}-\frac{n}{4(n-1)}\mathcal{R}^{2}]+...
\notag \\
&&-\sum_{s=[(n+1)/2]}^{\infty }\frac{B_{s}}{[(\alpha ^{2}-1)(n-2+\alpha
^{2})]^{s}}\left( \frac{le^{-\alpha \Phi }}{2}\right) ^{2s-1}\mathcal{R}%
^{s}\},  \label{Actct}
\end{eqnarray}%
where $[(n+1)/2]$ denotes the integer part of $(n+1)/2$. Again although each
term in the summation in Eq. (\ref{Actct}) may not be zero, the summation of
all the divergent terms in the total action is zero at infinity provided one
chooses $B_{s}$ correctly. Thus in order to calculate the finite action, one
only needs to consider the finite terms in the counterterms which are the
generalization of counterterms of Einstein gravity. That is, the finite
action is%
\begin{eqnarray}
I_{total} &=&\text{Finite terms of}\left\{ I_{bulk}+I_{GH}+\frac{1}{8\pi }%
\int d^{3}x\sqrt{-\gamma }\left[ \frac{2}{le^{-\alpha \Phi }}+\frac{%
le^{-\alpha \Phi }}{2(1-\alpha ^{4})}\mathcal{R}\right. \right.   \notag \\
&&\left. \left. +\frac{l^{3}e^{-3\alpha \Phi }}{2(n-4)(1-\alpha
^{2})(n-2+\alpha ^{2})^{2}}\left( \mathcal{R}_{ab}\mathcal{R}^{ab}-\frac{n}{%
4(n-1)}\mathcal{R}^{2}\right) +...\right] \right\} .  \label{Itot}
\end{eqnarray}%
The finite stress energy tensor and conserved charges are
\begin{eqnarray}
T^{ab} &=&\frac{1}{8\pi }\left\{ (K^{ab}-K\gamma ^{ab})+\frac{n-1}{%
le^{-\alpha \Phi }}\gamma ^{ab}-\frac{le^{-\alpha \Phi }}{(1-\alpha
^{2})(n-2+\alpha ^{2})}(\mathcal{R}^{ab}-\frac{1}{2}\mathcal{R}\gamma
^{ab})\right.   \notag \\
&&\ \ \left. -\frac{l^{3}e^{-3\alpha \Phi }}{(1-\alpha ^{2})(n-4)(n-2+\alpha
^{2})^{2}}[-\frac{1}{2}\gamma ^{ab}(\mathcal{R}^{cd}\mathcal{R}_{cd}-\frac{n%
}{4(n-1)}\mathcal{R}^{2})-\frac{n}{2(n-1)}\mathcal{RR}^{ab}\right.   \notag
\\
&&\ \ \left. +2\mathcal{R}_{cd}\mathcal{R}^{acbd}-\frac{n-2}{2(n-1)}\nabla
^{a}\nabla ^{b}\mathcal{R}+\nabla ^{2}\mathcal{R}^{ab}-\frac{1}{2(n-1)}%
\gamma ^{ab}\nabla ^{2}\mathcal{R}]+...\right.   \notag \\
&&\left. +\sum_{s=[(n+1)/2]}2sB_{s}\left( \frac{le^{-\alpha \Phi }}{2}%
\right) ^{2s-1}\left[ \mathcal{R}^{s-1}(\mathcal{R}^{ab}-\frac{1}{2s}%
\mathcal{R}\gamma ^{ab})-(\nabla ^{a}\nabla ^{b}-g^{ab}\nabla ^{2})\mathcal{R%
}^{s-1}\right] \right\} ,  \label{Str}
\end{eqnarray}%
\begin{equation}
\mathcal{Q}(\xi )=\int_{\Sigma }d^{n-1}x\sqrt{\sigma }u^{a}T_{ab}\xi ^{b},
\label{Cons}
\end{equation}%
respectively. Again, the summation of all the divergent terms in Eq. (\ref%
{Cons}) is zero at infinity and one should consider only the finite terms of
Eq. (\ref{Str}). In other words, since all the terms in the summation of Eq.
(\ref{Str}) are zero at infinity, one should consider only the counterterms
which are the generalization of counterterms of Einstein gravity in the
presence of dilaton without the summation term.

\section{Review of topological dilaton black holes with power-Maxwell field}

\label{PLM} We consider the action of $(n+1)$-dimensional $(n\geq 3)$
Einstein gravity with power-Maxwell Lagrangian which is coupled to the
dilaton field \cite{Kord}
\begin{equation}
I=-\frac{1}{16\pi }\int d^{n+1}x\sqrt{-g}\left\{ \mathcal{R}-\frac{4}{n-1}%
(\nabla \Phi )^{2}-V(\Phi )+\left( -e^{-4\alpha \Phi /(n-1)}F\right)
^{p}\right\} ,  \label{Act}
\end{equation}%
where $\mathcal{R}$ is the Ricci scalar, $\Phi $ is the dilaton field,$%
V(\Phi )$ is a potential for $\Phi $, and $p$ and $\alpha $ are two
constants determining the nonlinearity of the electromagnetic field and the
strength of coupling of the scalar and electromagnetic field, respectively. $%
F=F_{\lambda \eta }F^{\lambda \eta }$, where $F_{\mu \nu }=\partial _{\mu
}A_{\nu }-\partial _{\nu }A_{\mu }$ is the electromagnetic field tensor and $%
A_{\mu }$ is the electromagnetic potential. In order to obtain the field
equations, we vary the action (\ref{Act}) with respect to the gravitational
field $g_{\mu \nu }$, the dilaton field $\Phi $ and the gauge field $A_{\mu }
$. We find \cite{Kord}
\begin{eqnarray}
&&\mathcal{R}_{\mu \nu }=\left\{ \frac{1}{n-1}V(\Phi )+\frac{(2p-1)}{n-1}%
\left( -Fe^{-4\alpha \Phi /(n-1)}\right) ^{p}\right\} g_{\mu \nu }  \notag \\
&&+\frac{4}{n-1}\partial _{\mu }\Phi \partial _{\nu }\Phi +2pe^{-4\alpha
p\Phi /(n-1)}(-F)^{p-1}F_{\mu \lambda }F_{\nu }^{\text{ \ }\lambda },
\label{FE1}
\end{eqnarray}%
\begin{eqnarray}
\nabla ^{2}\Phi -\frac{n-1}{8}\frac{\partial V}{\partial \Phi }-\frac{%
p\alpha }{2}e^{-{4\alpha p\Phi }/({n-1})}(-F)^{p} &=&0,  \label{FE2} \\
\partial _{\mu }\left( \sqrt{-g}e^{-{4\alpha p\Phi }/({n-1}%
)}(-F)^{p-1}F^{\mu \nu }\right)  &=&0.  \label{FE3}
\end{eqnarray}%
In order to construct static topological black hole solutions of the above
field equations, we take the line elements of spacetime in the form
\begin{equation}
ds^{2}=-f(r)dt^{2}+{\frac{dr^{2}}{f(r)}}+r^{2}R^{2}(r)h_{ij}dx^{i}dx^{j},
\label{metric}
\end{equation}%
where $f(r)$ and $R(r)$ are functions of $r$ which should be determined, and
$h_{ij}$ is a function of coordinates $x^{i}$ which spanned an $(n-1)$%
-dimensional hypersurface with constant scalar curvature $(n-1)(n-2)k$. The
constant $k=0,-1$, and $+1$ for flat, hyperbolic and spherical
hypersurfaces. The solution of Eq. (\ref{FE3}) is given
\begin{equation}
F_{tr}=\frac{qe{^{{\frac{4\alpha p\Phi \left( r\right) }{\left( n-1\right)
\left( 2p-1\right) }}}}}{\left( rR\right) ^{{\frac{n-1}{2p-1}}}},
\label{Ftr}
\end{equation}%
where $q$ is an integration constant related to charge of the black hole.
Substituting (\ref{metric}) and (\ref{Ftr}) in the field equations (\ref{FE1}%
) and (\ref{FE2}), we arrive at
\begin{equation}
f^{\prime \prime }+{\frac{\left( n-1\right) f^{\prime }}{r}}+{\frac{\left(
n-1\right) f^{\prime }R^{\prime }}{R}}+{\frac{2V}{n-1}}-\frac{2[1+\left(
n-3\right) ]p}{n-1}\left( 2{q}^{2}\left( rR\right) ^{-{\frac{2(n-1)}{2p-1}}}e%
{^{{\frac{4\alpha \Phi }{\left( n-1\right) \left( 2p-1\right) }}}}\right)
^{p}=0,  \label{fe1}
\end{equation}

\begin{gather}
f^{\prime \prime }+{\frac{\left( n-1\right) f^{\prime }}{r}}+{\frac{\left(
n-1\right) f^{\prime }R^{\prime }}{R}}+{\frac{2V}{n-1}} +{\frac{4\left(
n-1\right) fR^{\prime }}{rR}}+{\frac{2\left( n-1\right) fR^{\prime \prime }}{%
R}}+{\frac{8f\Phi ^{\prime 2}}{n-1}}  \notag \\
-\frac{2[1+\left( n-3\right) p]}{n-1}\left( 2{q}^{2}\left( rR\right) ^{-{%
\frac{2(n-1)}{2p-1}}}e{^{{\frac{4\alpha \Phi }{\left( n-1\right) \left(
2p-1\right) }}}}\right) ^{p}=0,  \label{fe2}
\end{gather}

\begin{gather}
{\frac{f^{\prime }}{r}}+{\frac{f^{\prime }R^{\prime }}{R}+\frac{\left(
n-2\right) f}{{r}^{2}}+\frac{2\left( n-1\right) fR^{\prime }}{rR}} {+\frac{%
\left( n-2\right) R^{\prime 2}f}{R^{2}}+\frac{fR^{\prime \prime }}{R}}-{%
\frac{k\left( n-2\right) }{\left( rR\right) ^{2}}+\frac{V}{n-1}}  \notag \\
+\frac{2p-1}{n-1}\left( 2{q}^{2}\left( rR\right) ^{-{\frac{2(n-1)}{2p-1}}}e{%
^{{\frac{4\alpha \Phi }{\left( n-1\right) \left( 2p-1\right) }}}}\right)
^{p}=0,  \label{fe3}
\end{gather}

\begin{gather}
f\Phi ^{\prime \prime }+\Phi ^{\prime }f^{\prime }+{\frac{\left( n-1\right)
f\Phi ^{\prime }}{r}} +{\frac{\left( n-1\right) f\Phi ^{\prime }R^{\prime }}{%
R}}-\frac{n-1}{8}\frac{dV}{d\Phi }  \notag \\
-\frac{p\alpha }{2}\left( 2{q}^{2}\left( rR\right) ^{-{\frac{2(n-1)}{2p-1}}}e%
{^{{\frac{4\alpha \Phi }{\left( n-1\right) \left( 2p-1\right) }}}}\right)
^{p}=0,  \label{fe4}
\end{gather}%
where the prime stands for the derivative with respect to $r$. It was argued
in \cite{Kord} that in order to have exact topological solutions with an
arbitrary dilaton coupling parameter $\alpha $, the dilaton potential should
be chosen with the combination of three Liouville-type,
\begin{equation}
V(\Phi )=2\Lambda _{1}e{^{2\zeta _{1}\Phi }}+2\Lambda _{2}e{^{2\zeta
_{2}\Phi }}+2\Lambda e{^{2\zeta _{3}\Phi }},  \label{v2}
\end{equation}%
where $\Lambda _{1}$, $\Lambda _{2}$, $\Lambda $, $\zeta _{1}$, $\zeta _{2}$
and $\zeta _{3}$ are constants. Note that the topological black holes in EMd
theory, can be constructed with two Liouville terms in the dilaton potential
\cite{CHM,Shey}.

In order to solve the system of equations (\ref{fe1})-(\ref{fe4}) for three
unknown functions $f(r)$, $R(r)$ and $\Phi (r)$, we make the ansatz
\begin{equation}
R(r)=e{^{{{2\alpha \Phi \left( r\right) }/({n-1})}}}.  \label{Rphi}
\end{equation}%
Subtracting (\ref{fe1}) from (\ref{fe2}), after using (\ref{Rphi}), we find
the following equation for the scalar field,
\begin{equation}
\Phi ^{\prime \prime }+\frac{2(\alpha ^{2}+1)\Phi ^{\prime 2}}{\alpha (n-1)}+%
\frac{2\Phi ^{\prime }}{r}=0,
\end{equation}%
which admits the following solution
\begin{equation}
\Phi (r)=\frac{\left( n-1\right) \alpha }{2\left( {\alpha }^{2}+1\right) }%
\ln \left( {\frac{b}{r}}\right) .  \label{phi}
\end{equation}%
Substituting (\ref{Rphi}) and (\ref{phi}) in Eqs. (\ref{fe2})-(\ref{fe4}),
one can easily show that these equation have a unique consistent solution of
the form \cite{Kord}
\begin{eqnarray}
f(r) &=&\frac{k\left( n-2\right) (1+\alpha ^{2})^{2}{r}^{2\gamma }}{%
(1-\alpha ^{2})\left( {\alpha }^{2}+n-2\right) {b}^{2\gamma }}-\frac{m}{{r}%
^{(n-1)(1-\gamma )-1}}+\frac{2^{p}p(1+\alpha ^{2})^{2}\left( 2p-1\right) ^{2}%
{b}^{-{\frac{2\left( n-2\right) p\gamma }{\left( 2p-1\right) }}}{q}^{2p}}{%
\Pi \left( n+{\alpha }^{2}-2p\right) {r}^{-\frac{2[\left( n-3\right)
p+1]-2p\left( n-2\right) \gamma }{2p-1}}}  \notag \\
&&-\frac{2\Lambda {b}^{2\gamma }(1+\alpha ^{2})^{2}{r}^{2(1-\gamma )}}{%
\left( n-1\right) \left( n-{\alpha }^{2}\right) },  \label{fr}
\end{eqnarray}%
where $b$ is an arbitrary non-zero positive constant, $\gamma =\alpha
^{2}/(\alpha ^{2}+1)$, $\Pi ={\alpha }^{2}+\left( n-1-{\alpha }^{2}\right) p$%
, and the constants should be fixed as
\begin{gather}
\zeta _{1}={\frac{2}{\left( n-1\right) \alpha }},\hspace{0.8cm}\zeta _{2}={%
\frac{2p\left( n-1+{\alpha }^{2}\right) }{\left( n-1\right) \left(
2p-1\right) \alpha }},  \notag \\
\zeta _{3}={\frac{2\alpha }{n-1}},\hspace{0.8cm}\Lambda _{1}={\frac{k\left(
n-1\right) \left( n-2\right) {\alpha }^{2}}{2{b}^{2}\left( {\alpha }%
^{2}-1\right) }},  \notag \\
\Lambda _{2}=\frac{2^{p-1}\left( 2p-1\right) \left( p-1\right) {\alpha }^{2}{%
q}^{2p}}{\Pi {b}^{{\frac{2\left( n-1\right) p}{2p-1}}}}.  \label{lam0}
\end{gather}

It is worth noting that in the linear Maxwell case where $p=1$, we have $%
\Lambda _{2}=0$ and hence the potential has two terms. Indeed, the term $%
2\Lambda _{2}e{^{2\zeta _{2}\Phi }}$ in the Liouville potential is necessary
in order to have solution (\ref{fr}) for the field equations of power-law
Maxwell field in dilaton gravity. Note that $\Lambda $ remains as a free
parameter which plays the role of the cosmological constant and we assume to
be negative and take it in the form $\Lambda =-(n-\alpha ^{2})(n-1)/2l^{2}$.
The parameter $m$ in Eq. (\ref{fr}) is the integration constant which is
known as the geometrical mass and can be written in term of horizon radius as%
\begin{eqnarray}
m(r_{+}) &=&\frac{k\left( n-2\right) {b}^{-2\gamma }{r}_{+}^{{\frac{{\alpha }%
^{2}+n-2}{{\alpha }^{2}+1}}}}{(2\gamma -1)(\gamma -1)\left( {\alpha }%
^{2}+n-2\right) }+\frac{2^{p}p\left( 2p-1\right) ^{2}{b}^{{\frac{-2\left(
n-2\right) \gamma p}{\left( 2p-1\right) }}}{q}^{2p}{r}_{+}^{-{\frac{{\alpha }%
^{2}-2p+n}{\left( 2p-1\right) \left( {\alpha }^{2}+1\right) }}}}{(\gamma
-1)^{2}\left( {\alpha }^{2}-2p+n\right) \Pi }  \notag \\
&&+\frac{{b}^{2\gamma }n{r}_{+}^{-{\frac{{\alpha }^{2}-n}{{\alpha }^{2}+1}}}%
}{l^{2}(\gamma -1)^{2}\left( n-{\alpha }^{2}\right) },  \label{mrh}
\end{eqnarray}%
where $r_{+}$ is the positive real root of $f(r_{+})=0$. In the limiting
case where $p=1$, solution (\ref{fr}) reduces to the topological dilaton
black holes of EMd gravity presented in Ref. \cite{CHM,Shey}. In case of
linear Maxwell theory ($p=1$) and in the absence of dilaton field ($\alpha
=\gamma =0$), solution (\ref{fr}) reduces to asymptotically AdS topological
black hole (see for example \cite{Cai3})
\begin{equation}
f(r)=k-\frac{m}{r^{n-2}}+\frac{2q^{2}}{(n-1)(n-2)r^{2(n-2)}}-\frac{2\Lambda
}{n(n-1)}r^{2}.
\end{equation}%
The gauge potential $A_{t}$ corresponding to the electromagnetic field (\ref%
{Ftr}) is given by
\begin{equation}
A_{t}=\frac{q{b}^{{\frac{\left( 2p+1-n\right) \gamma }{\left( 2p-1\right) }}}%
}{\Upsilon {r}^{\Upsilon }},  \label{At}
\end{equation}%
where $\Upsilon ={(n-2p+\alpha }^{2}{)/[(2p-1)(1+\alpha }^{2})]$. Imposing
two conditions, namely (i) the electric potential $A_{t}$ should have a
finite value at infinity and (ii) the term including $m$ in spacial infinity
should vanish, lead to the following restrictions on the parameters $p$ and $%
\alpha $ \cite{Kord},
\begin{eqnarray}
\text{For }\frac{1}{2} &<&p<\frac{n}{2}\text{, \ \ \ \ \ \ \ \ }0\leq \alpha
^{2}<n-2,  \label{res1} \\
\text{For }\frac{n}{2} &<&p<n-1\text{, \ \ \ \ }2p-n<\alpha ^{2}<n-2.
\label{res2}
\end{eqnarray}%
The Hawking temperature can be written as

\begin{equation}
T_{+}=\frac{(1+\alpha ^{2})}{4\pi }\left( \frac{k(n-2)}{b^{2\gamma
}(1-\alpha ^{2})r_{+}^{1-2\gamma }}-\frac{2\Lambda b^{2\gamma
}r_{+}^{1-2\gamma }}{n-1}-\frac{2^{p}p\left( 2p-1\right) {b}^{{\frac{%
-2\left( n-2\right) \gamma p}{\left( 2p-1\right) }}}{q}^{2p}}{\Pi {r}_{+}^{%
\frac{2p(n-2)(1-\gamma )+1}{2p-1}}}\right) .  \label{Temp}
\end{equation}%
One can calculate thermodynamic quantities such as entropy, charge and
electric potential of the black hole per unit volume ${\omega _{n-1}}$ as
\cite{Kord}%
\begin{equation}
{S}=\frac{b^{(n-1)\gamma }r_{+}^{(n-1)(1-\gamma )}}{4}.  \label{Entropy}
\end{equation}%
\begin{equation}
Q=\frac{\tilde{q}}{4\pi },\ \ \ \tilde{q}={2^{p-1}{q}^{2p-1}}.
\label{Charge}
\end{equation}%
\begin{equation}
U=\frac{Cq{b}^{{\frac{\left( 2p-n+1\right) \gamma }{\left( 2p-1\right) }}}}{%
\Upsilon {r}_{+}^{\Upsilon }}.  \label{Pot1}
\end{equation}%
where $C=\left( n-1\right) {p}^{2}/\Pi $. For $p=1$ we have $\tilde{q}=q$,
as expected.

\section{Extended Phase space and Smarr Formula \label{Ph}}

In this section, we would like to investigate the thermodynamics of
topological dilaton black holes with power-Maxwell field. First, we
calculate the mass through the use of \ counterterm method. Using Eqs. (\ref%
{Stress}) and (\ref{Cons}), one obtains%
\begin{equation}
{M}=\frac{b^{(n-1)\gamma }(n-1)}{16\pi (\alpha ^{2}+1)}m.
\end{equation}%
We shall construct a Smarr relation in an extended phase space in which the
cosmological constant is treated as thermodynamic variable. The conjugate
quantity of the cosmological constant, which is proportional to the
pressure, is the volume. As we know the entropy of black hole is a quarter
of the area of the horizon, so the thermodynamic volume $V$ is obtained as
\begin{equation}
V=\int 4Sdr_{+}=\frac{\alpha ^{2}+1}{\alpha ^{2}+n}b^{(n-1)\gamma }r_{+}^{%
\frac{\alpha ^{2}+n}{\alpha ^{2}+1}}\omega _{n-1}.  \label{volume}
\end{equation}%
In an extended phase space, $M$ can be a function of thermodynamic
quantities entropy, pressure and charge. Hence, the first law takes the form
\begin{equation}
dM=TdS+UdQ+PdV.
\end{equation}%
It is easy to show that the conjugate quantity of the thermodynamic volume
is \cite{Kamrani}
\begin{equation}
P=-\frac{\Lambda }{8\pi }\frac{n-\gamma (n-1)}{n-\gamma (n+1)}\left( \frac{b%
}{r_{+}}\right) ^{2\gamma }=-\frac{\left( n+\alpha ^{2}\right) }{8\pi \left(
n-\alpha ^{2}\right) }\left( \frac{b}{r_{+}}\right) ^{2\gamma }\Lambda ,
\label{press}
\end{equation}%
which is proportional to the cosmological constant $\Lambda $. In the
absence of dilaton ($\gamma =0=\alpha $) we can see that the above
expression for $P$ reduces to the pressure of the RN-AdS black holes \cite%
{MannRN}. One may note that the above expression for the pressure is the
same as that of EMd \cite{Kamrani} and EBId black holes \cite{dayyani}.
Also, it is clear that the pressure is positive provided $\alpha <\sqrt{n}$
exactly similar to \cite{Kamrani} and \cite{dayyani}. This is consistent
with the argument given in \cite{Kord}, which states that the topological
dilaton black hole solutions have reasonable behavior provided $\alpha <%
\sqrt{n-2}$. The Smarr relation may be obtained from the first law of black
hole thermodynamics and a scaling dimensional argument \cite{Ka}. One
obtains
\begin{equation}
M=\frac{n-1}{n-2+\alpha ^{2}}TS+\frac{p(n-3+\alpha ^{2})+1}{p(n-2+\alpha
^{2})}UQ+\frac{2(\alpha ^{2}-1)}{n-2+\alpha ^{2}}VP.
\end{equation}%
One may note that the above generalized Smarr formula reduces to those of
Refs. \cite{Kamrani} in the limit $p=1$.

In what follows, we study the phase transition of the power-Maxwell dilaton
black holes in an extended phase space in both canonical and grand-canonical
ensembles, separately.

\section{Phase Transition in Canonical Ensemble}

\label{Can} In order to study the phase transition of system, we first
consider the canonical ensemble. In this ensemble the charge $Q$ of the
black hole is regarded as a fixed extensive parameter.

\subsection{Equation of state}

\begin{figure}[tbp]
\epsfxsize=5cm \centerline{\epsffile{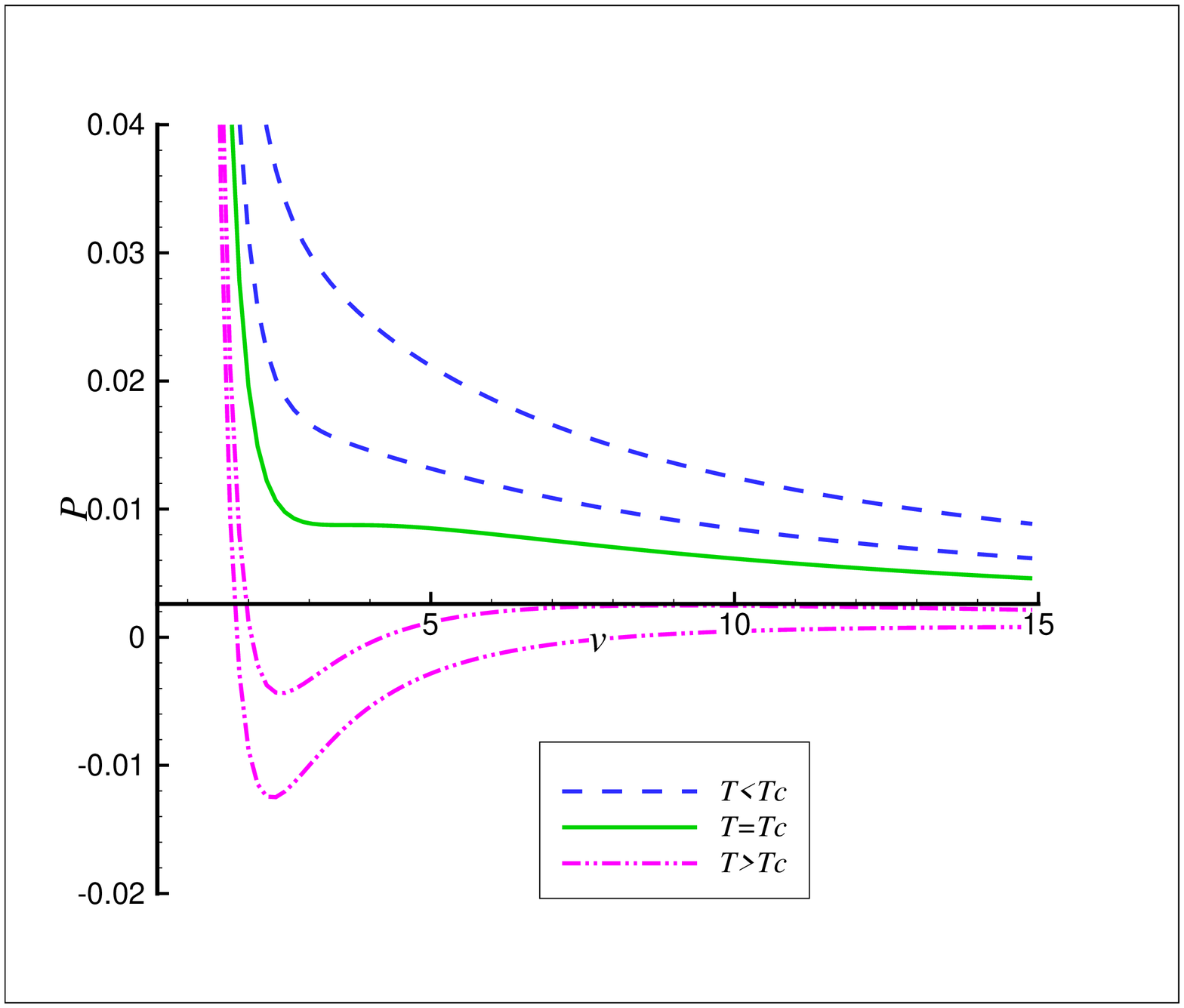}}
\caption{$P-v$ diagram for power-Maxwell dilaton black holes. Here we have
taken $b=1$, $q=1$, $n=3$, $k=1$, $p=0.8$ and $\protect\alpha =0.3$.}
\label{Fig1}
\end{figure}

\begin{figure}[tbp]
\epsfxsize=5cm \centerline{\epsffile{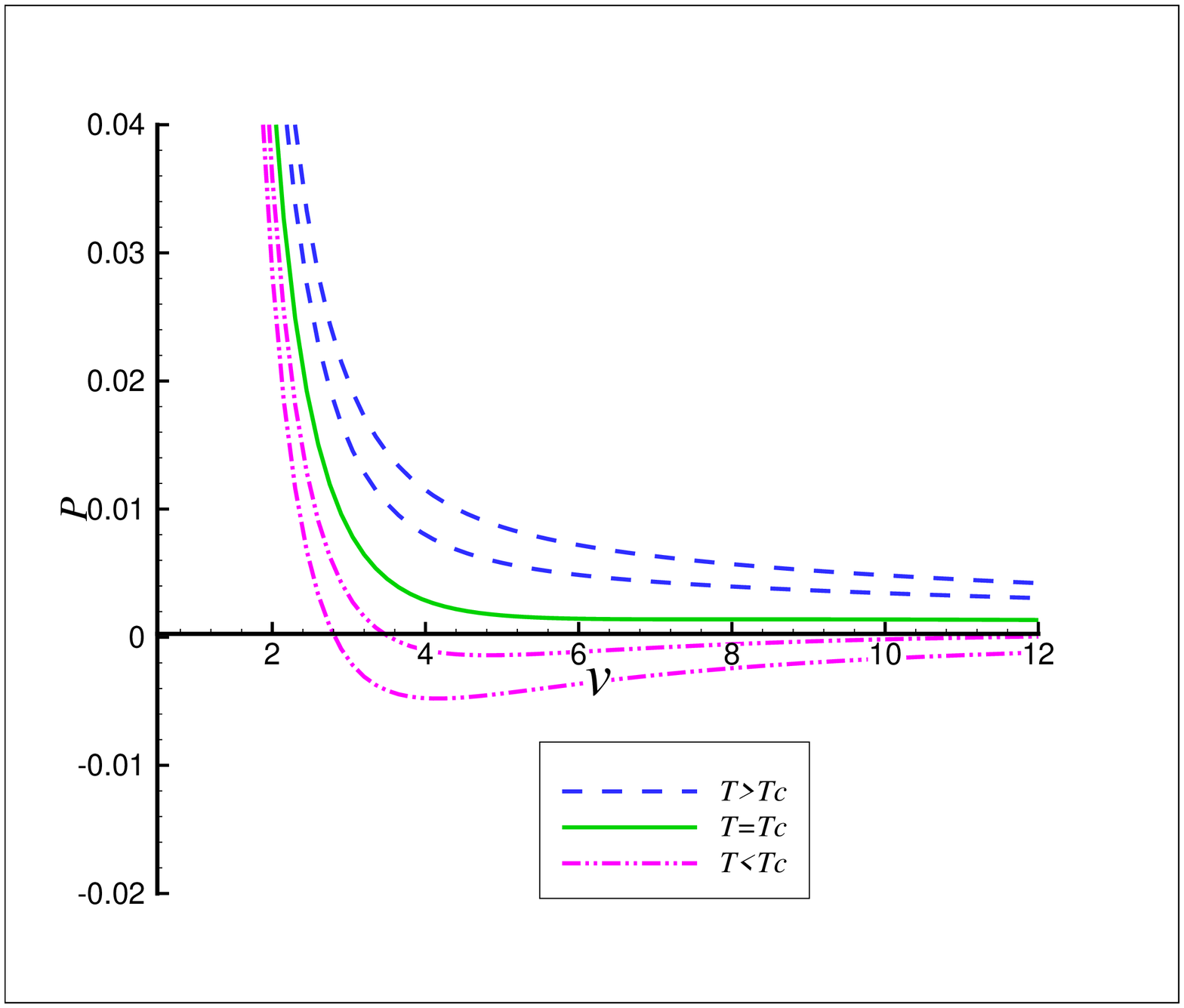}}
\caption{$P-v$ diagram for power-Maxwell dilaton black holes. Here we have
taken $b=1$, $q=1$, $n=3$, $k=1$, $p=1.2$ and $\protect\alpha =0.3$.}
\label{Fig2}
\end{figure}

\begin{figure}[tbp]
\epsfxsize=5cm \centerline{\epsffile{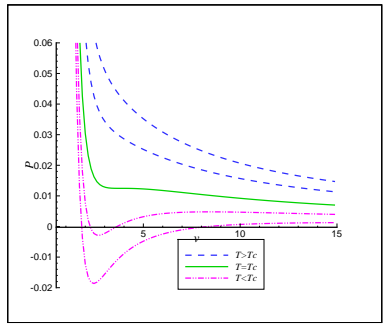}}
\caption{$P-v$ diagram for power-Maxwell dilaton black holes. Here we have
taken $b=1$, $q=1$, $n=5$, $k=1$, $p=3$ and $\protect\alpha =0.3$.}
\label{Fig3}
\end{figure}

\begin{figure}[tbp]
\epsfxsize=5cm \centerline{\epsffile{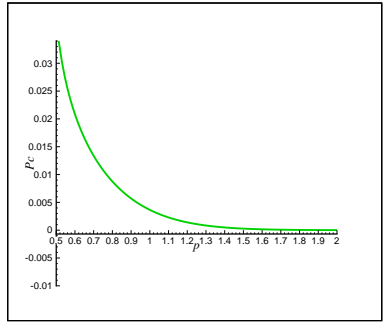}}
\caption{The behavior of $P_{c}$ versus $p$ for $b=1$, $q=1$, $n=3$, $k=1$
and $\protect\alpha =0.3$.}
\label{Fig11}
\end{figure}

\begin{figure}[tbp]
\epsfxsize=5cm \centerline{\epsffile{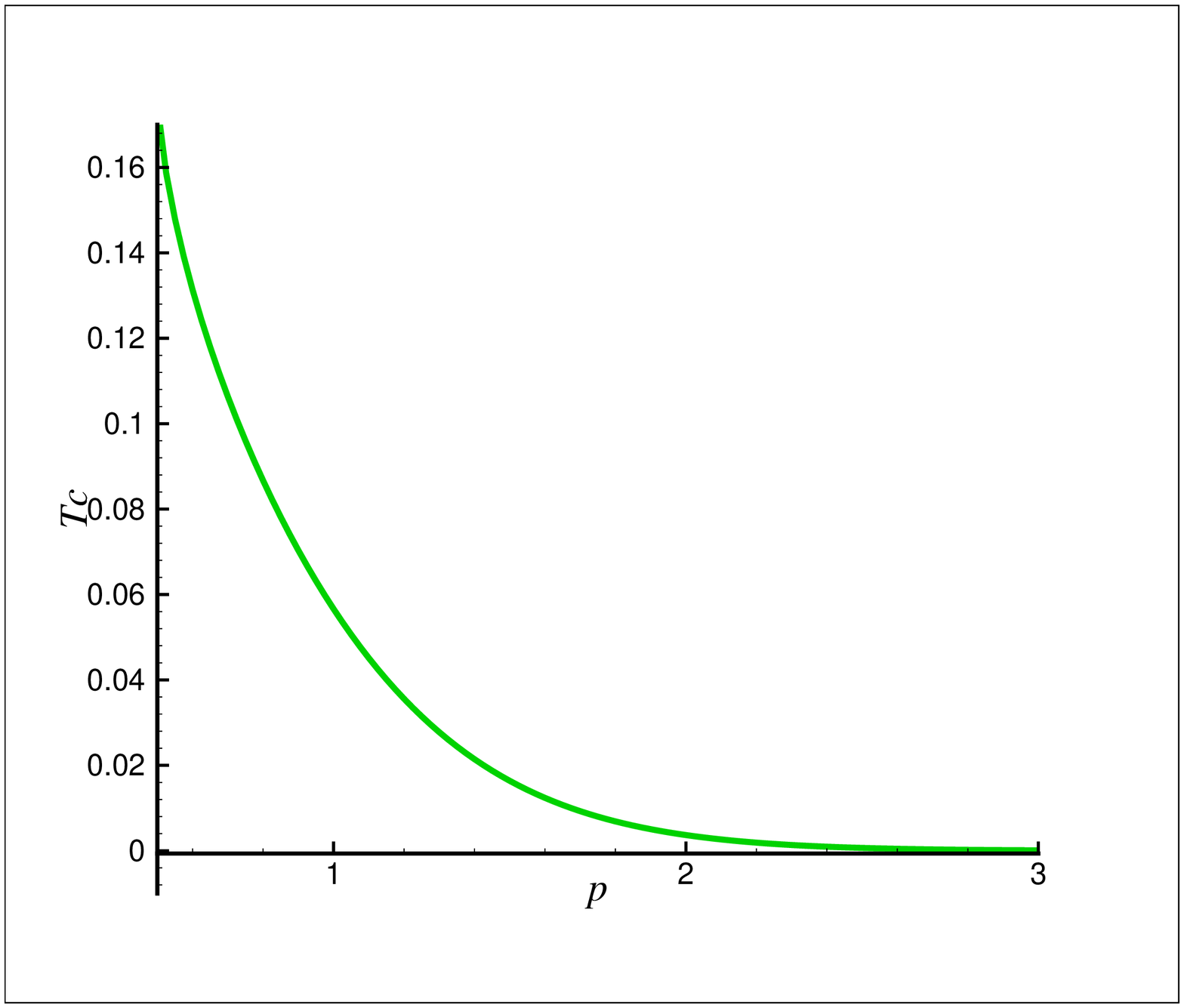}}
\caption{The behavior of $T_{c}$ versus $p$ for $b=1$, $q=1$, $n=3$, $k=1$
and $\protect\alpha =0.3$.}
\label{Fig12}
\end{figure}

Using Eqs. (\ref{Temp}) and (\ref{press}) for fixed charge, one may write
\begin{equation}
P=\frac{\Gamma T}{r_{+}}-\frac{k(n-2)(1+\alpha ^{2})\Gamma }{4\pi (1-\alpha
^{2})b^{2\gamma }r_{+}^{2-2\gamma }}+\frac{2^{p}p(1+\alpha ^{2})\left(
2p-1\right) {b}^{-{\frac{2\left( n-2\right) p\gamma }{\left( 2p-1\right) }}%
}\Gamma {q}^{2p}}{4\pi \Pi {r_{+}}^{\frac{2p(n-n\gamma -1+2\gamma )}{2p-1}}},
\label{eq of state1}
\end{equation}%
where
\begin{equation}
\Gamma =\frac{\left( n-1\right) \left( n+\alpha ^{2}\right) }{4\left(
n-\alpha ^{2}\right) \left( \alpha ^{2}+1\right) }.
\end{equation}%
From Eq. (\ref{volume}), we see that $r_{+}$ is a function of the
thermodynamic volume $V$, so the above equation can be regarded as the
equation of state $P(V,T)$. Before proceeding further, we translate the
`geometric' equation of state (\ref{eq of state1}) to a physical case by
performing a dimensional analysis. We identify the following relations
between geometric quantities and physical pressure and temperature
\begin{equation}
\mathcal{P}=\frac{\hbar c}{l_{p}^{2}}P,\quad \mathcal{T}=\frac{\hbar c}{%
\kappa }T,
\end{equation}%
where the Planck length is $l_{p}=\sqrt{\hbar G/c^{3}}$ and $\kappa $ is the
Boltzmann constant. In terms of these new definitions, Eq. (\ref{eq of
state1}) can be written as
\begin{equation*}
\mathcal{P}=\frac{\Gamma \kappa \mathcal{T}}{l_{p}^{2}r_{+}}-\frac{k\hbar
c(n-2)(1+\alpha ^{2})\Gamma }{4\pi l_{p}^{2}(1-\alpha ^{2})b^{2\gamma
}r_{+}^{2-2\gamma }}+\frac{2^{p}p(1+\alpha ^{2})\left( 2p-1\right) {b}^{-{%
\frac{2\left( n-2\right) p\gamma }{\left( 2p-1\right) }}}\Gamma {q}%
^{2p}\hbar c}{4\Pi \pi l_{p}^{2}{r_{+}}^{\frac{2p(n-n\gamma -1+2\gamma )}{%
2p-1}}}.
\end{equation*}%
Now, comparing the above physical equation of state with the van der Walls
equation \cite{MannRN}%
\begin{equation*}
\mathcal{P}=\frac{\mathcal{T}}{v}+...,
\end{equation*}%
we understand that the specific volume $v$ of the fluid in terms of the
horizon radius should be written as,
\begin{equation}
v=\frac{l_{p}^{2}r_{+}}{\Gamma },  \label{volume2}
\end{equation}%
Returning to the geometrical units $(G=\hbar =c=1\Longrightarrow
l_{p}^{2}=1) $, the equation of state (\ref{eq of state1}) can be written
\begin{equation}
P=\frac{T}{v}-\frac{k(n-2)(1+\alpha ^{2})\Gamma }{4\pi (1-\alpha
^{2})b^{2\gamma }(\Gamma v)^{2-2\gamma }}+\frac{2^{p}p(1+\alpha ^{2})\left(
2p-1\right) {b}^{-{\frac{2\left( n-2\right) p\gamma }{\left( 2p-1\right) }}%
}\Gamma {q}^{2p}}{4\Pi \pi \left( \Gamma v\right) ^{\frac{2p(n-n\gamma
-1+2\gamma )}{2p-1}}}.  \label{PvT}
\end{equation}%
To compare the critical behavior of the system with van der Waals fluid, we
should plot isotherm diagrams. The corresponding $P-v$ diagrams are
displayed in Figs. \ref{Fig1}-\ref{Fig3}.
The behavior of the isotherms diagrams depends on how deep we are in
nonlinear regime and the value of $n$. We observe that in order to have
critical behavior, the dimension of spacetime, $n$, should increase with
increasing the power parameter $p$. This can be easily understand as
follows. The second term in (\ref{PvT}) is independent of $p$ while the
third term depends on $p$ sensitively. To see critical behavior, the power
of $v$ in the denominator of the third term should be larger than this power
in the second term. From the other point of view, to satisfy (\ref{res1}), $%
n $ should increase as $p$ increases. The critical point can be obtained by
solving the following equations
\begin{equation}
\frac{\partial P}{\partial v}\Big|_{T_{c}}=0,\quad \frac{\partial ^{2}P}{%
\partial v^{2}}\Big|_{T_{c}}=0,
\end{equation}%
which leads to
\begin{eqnarray}
v_{c} &=&\frac{1}{\Gamma }X^{\frac{(2p-1)(\alpha ^{2}+1)}{2\Delta }}b^{\frac{%
-\alpha ^{2}(pn-4p+1)}{\Delta }}, \\
P_{c} &=&\frac{k(n-2)\Gamma \Delta }{4\pi p(n+\alpha ^{2}-1)}X^{\frac{-(2p-1)%
}{\Delta }}b^{\frac{-2p\alpha ^{2}}{\Delta }}, \\
T_{c} &=&\frac{k\Delta (n-2)}{\pi (1-\alpha ^{2})(2pn+\alpha ^{2}+1-4p)}X^{%
\frac{(2p-1)(\alpha ^{2}-1)}{2\Delta }}b^{\frac{-\alpha ^{2}(pn-2p+1)}{%
\Delta }},
\end{eqnarray}%
where
\begin{equation*}
X=\frac{(-4p+\alpha ^{2}+1+2pn)(n+\alpha ^{2}-1)q^{2p}p^{2}2^{p}}{(2p-1)k\Pi
(n-2)},
\end{equation*}%
\begin{equation*}
\Delta =pn+p\alpha ^{2}-3p+1.
\end{equation*}%
Using the above critical values, $\rho _{c}$ is obtained as
\begin{equation}
\rho _{c}=-\frac{(-4p+\alpha ^{2}+1+2pn)(\alpha ^{2}-1)}{4p(n+\alpha ^{2}-1)}%
,  \label{universal ratio}
\end{equation}%
As one expects, the above $\rho _{c}$ reduces to that of Ref. \cite{Kamrani}
for $p=1$,
\begin{equation}
\rho _{c}=\frac{P_{c}v_{c}}{T_{c}}=\frac{(1-\alpha ^{2})(2n-3+\alpha ^{2})}{%
(4n-4+4\alpha ^{2})},
\end{equation}%
and in the absence of the dilaton field ($\alpha =0=\gamma $) in four
dimensions ($n=3$), it reduces to $3/8$ which is the characteristic of van
der Waals fluid \cite{MannRN}. It is easy to see that $\rho _{c}$ is
positive provided $\alpha <1$. It is also notable that $P_{c}$ and $T_{c}$
decrease as $p$ increases (see Figs. \ref{Fig11}-\ref{Fig12}).
\begin{figure}[tbp]
\epsfxsize=5cm \centerline{\epsffile{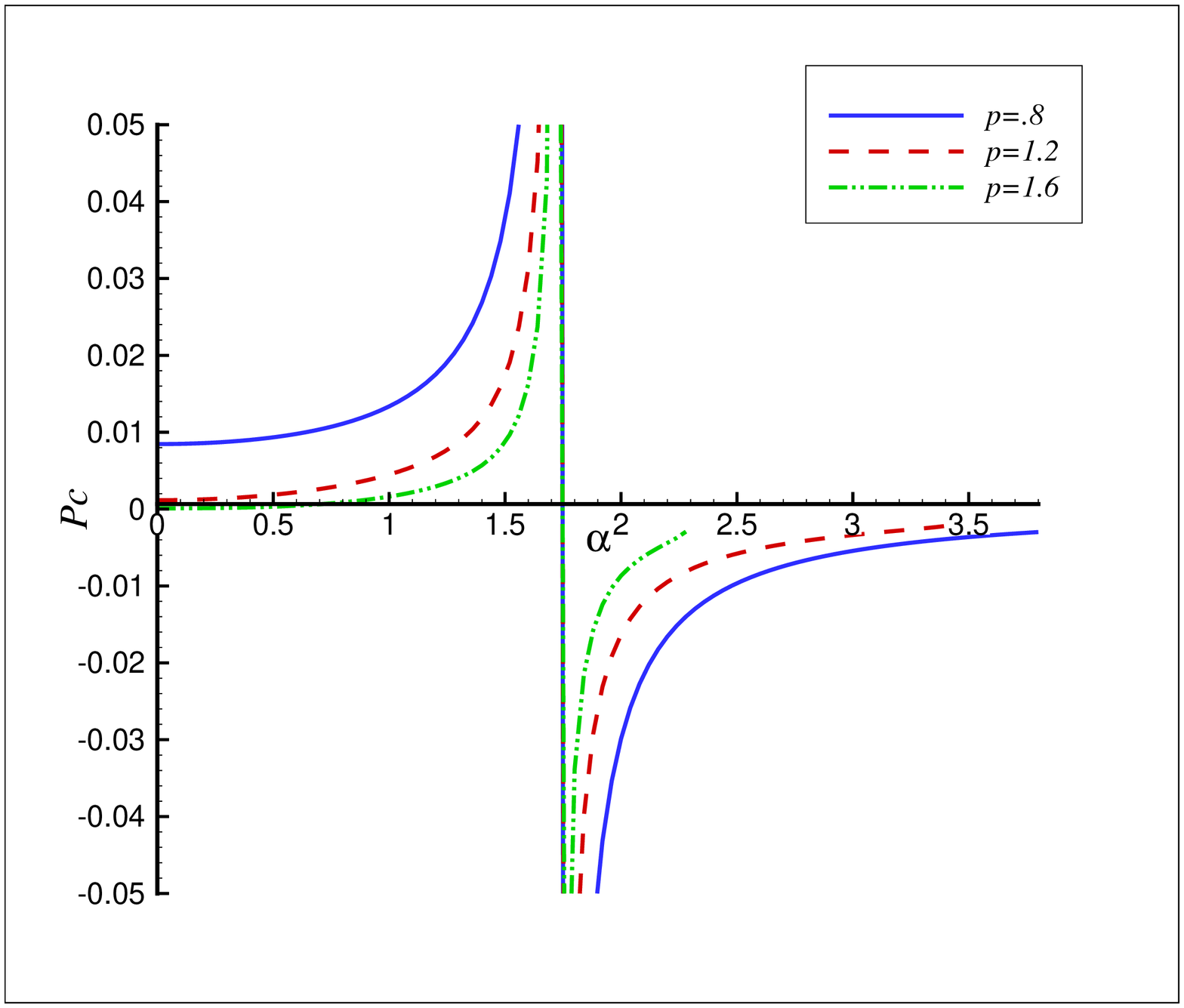}}
\caption{(Color online) $P_{c}-\protect\alpha $ diagram of power-Maxwell
dilaton black holes for $b=1$, $q=1$, $n=3$, $q=1$ and $k=1$. }
\label{Fig13}
\end{figure}
\begin{figure}[tbp]
\epsfxsize=5cm \centerline{\epsffile{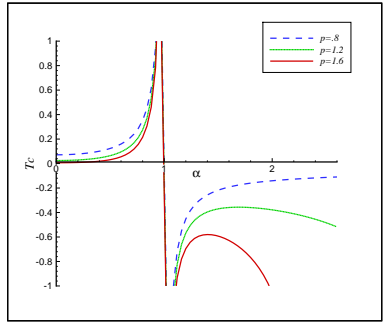}}
\caption{(Color online) $T_{c}-\protect\alpha $ diagram of power-Maxwell
dilaton black holes for $b=1$, $q=1$, $n=3$ and $k=1$. }
\label{Fig14}
\end{figure}

\subsection{Critical exponents}

The behavior of physical quantities in the vicinity of critical point can be
characterized by the critical exponents. So, following the approach of \cite%
{MannBI}, one can calculate the critical exponents $\alpha ^{\prime }$, $%
\beta ^{\prime }$, $\gamma ^{\prime }$ and $\delta ^{\prime }$ for the phase
transition of an $(n+1)$-dimensional charged dilaton black hole in the
presence of power-Maxwell field. To obtain the critical exponents, we define
the reduced thermodynamic variables as
\begin{equation*}
p=\frac{P}{P_{c}},\quad \nu =\frac{v}{v_{c}},\quad \tau =\frac{T}{T_{c}}.
\end{equation*}%
Therefore, equation of state (\ref{PvT}) translate into the law of
corresponding state,
\begin{eqnarray}
p &=& \frac{1}{\rho_{c}} \frac{\tau}{\nu} -\frac{k(n-2)(1+\alpha ^{2})\Gamma
}{4\pi P_{c} (1-\alpha ^{2})b^{2\gamma }(\Gamma \nu v_{c})^{2-2\gamma }} +%
\frac{2^{p}p(1+\alpha ^{2})\left( 2p-1\right) {b}^{-{\frac{2\left(
n-2\right) p\gamma }{\left( 2p-1\right) }}}\Gamma {q}^{2p}}{4\pi \Pi P_{c}
\left(\Gamma \nu v_{c}\right)^{\frac{2p(n-n\gamma-1+2\gamma) }{2p-1}}}.
\label{law}
\end{eqnarray}%
Although this law depends on parameter $\gamma $, and $p$ but as we will see
this doesn't affect the behavior of the critical exponents. To calculate the
critical exponent $\alpha ^{\prime }$, we consider the entropy $S$ (\ref%
{Entropy}) as a function of $T$ and $V$.\ Using (\ref{volume}) we have
\begin{equation}
S=S\left( T,V\right) =\frac{b^{(n-1)\gamma }\omega _{n-1}}{4}\left[V\left[
(n-1)(1-\gamma )+1\right] /(b^{(n-1)\gamma }\omega _{n-1})\right]^{\frac{%
(n-1)(1-\gamma )}{(n-1)(1-\gamma )+1}}.
\end{equation}
Obviously, this is independent of $T$ and therefore the specific heat
vanishes, $C_{V}=T\left( \partial S/\partial T\right) _{V}=0$. Since the
exponent $\alpha ^{\prime }$ governs the behavior of the specific heat at
constant volume $C_{V}\varpropto \left\vert \tau -1\right\vert ^{\alpha
^{\prime }}$, hence we have $\alpha ^{\prime }=0$. Expanding Eq. (\ref{law})
near the critical point
\begin{equation}
\tau =t+1,\quad \nu =\left( \omega +1\right) ^{\frac{1}{\varepsilon }},
\end{equation}%
where $\varepsilon $ is a positive parameter defined as $\varepsilon
=n-\gamma \left( n-1\right) =(n+\alpha ^{2})/(1+\alpha ^{2})$ and following
the method of Ref. \cite{MannBI}, we obtain
\begin{equation}
p=1+At-Bt\omega -C\omega ^{3}+O\left( t\omega ^{2},\omega ^{4}\right) ,
\label{ptw}
\end{equation}%
with%
\begin{eqnarray}
A &=&\frac{1}{\rho _{c}},\quad B=\frac{1}{\varepsilon \rho _{c}}, \\
C &=&\frac{2(n-1+\alpha ^{2})}{3(2p-1)(1+\alpha ^{2})^{2}\varepsilon ^{3}}.
\end{eqnarray}%
Differentiating Eq.\ (\ref{ptw}) at a fixed $t<0$ with respect to $\omega $,
we get
\begin{equation}
dP=-P_{c}\left( Bt+3C\omega ^{2}\right) d\omega .
\end{equation}%
Now, we apply the Maxwell's equal area law \cite{MannRN}. Denoting the
volume of small and large black holes with $\omega _{s}$ and $\omega _{l}$,
respectively, we obtain
\begin{eqnarray}
p &=&1+At-Bt\omega _{l}-C\omega _{l}^{3}=1+At-Bt\omega _{s}-C\omega _{s}^{3},
\notag \\
0 &=&\int_{\omega _{l}}^{\omega _{s}}\omega dP.  \label{Equal}
\end{eqnarray}%
Eq. (\ref{Equal}) leads to the unique non-trivial solution
\begin{equation}
\omega _{l}=-\omega _{s}=\sqrt{-\frac{Bt}{C}}.  \label{oml}
\end{equation}
which gives the order parameter $\eta =V_{c}\left( \omega _{l}-\omega
_{s}\right) $ as
\begin{equation}
\eta =2V_{c}\omega _{l}=2\sqrt{-\frac{B}{C}}t^{1/2}.
\end{equation}%
\newline
Thus, the exponent $\beta ^{\prime }$ which describes the behavior of the
order parameter $\eta $ near the critical point is $\beta ^{\prime }=1/2$.%
\newline
To calculate the exponent $\gamma ^{\prime }$, we may determine the behavior
of the isothermal compressibility near the critical point. Differentiating
Eq.(\ref{ptw}) with respect to $V$, one obtains
\begin{equation}
\frac{\partial V}{\partial P}\Big|_{T}=-\frac{V_{c}}{BP_{c}}\frac{1}{t}%
+O(\omega ).
\end{equation}
Hence, the isothermal compressibility near the critical point may be written
as
\begin{equation}
\kappa _{T}=-\frac{1}{V}\frac{\partial V}{\partial P}\Big|_{T}\propto -\frac{%
V_{c}}{BP_{c}}\frac{1}{t}\quad \Longrightarrow \quad \gamma ^{\prime }=1.
\end{equation}
Finally, the shape of the critical isotherm $t=0$ is given by (\ref{ptw})
\begin{equation}
p-1=-C\omega ^{3}\quad \Longrightarrow \quad \delta ^{\prime }=3.
\end{equation}
Thus, we have shown that for power-Maxwell-dilaton black holes in $(n+1)$
dimensions, the critical exponents have the same values as in case of the
RN-AdS black holes \cite{MannRN}, EMd black holes \cite{Kamrani} and EBId
black holes \cite{dayyani}.

\subsection{Gibbs free energy in canonical ensemble \label{Gibbs}}

\begin{figure}[tbp]
\epsfxsize=5cm \centerline{\epsffile{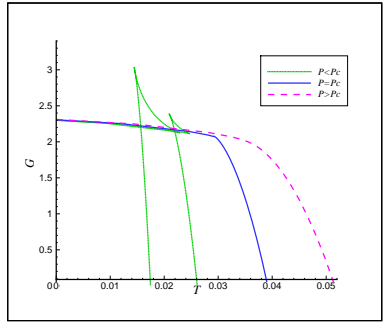}}
\caption{Gibbs free energy versus $T$ for $b=1$, $n=3$, $q=1$, $k=1$, $p
=1.2 $ and $\protect\alpha=0.2$.}
\label{Fig4}
\end{figure}

\begin{figure}[tbp]
\epsfxsize=5cm \centerline{\epsffile{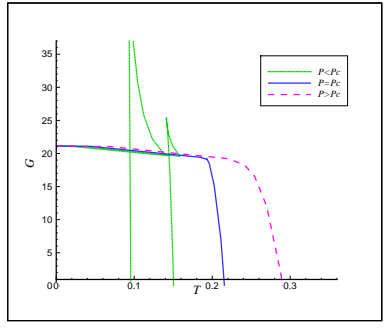}}
\caption{Gibbs free energy versus $T$ for $b=1$, $n=5$, $q=1$, $k=1$, $p= 2$
and $\protect\alpha=0.2$.}
\label{Fig5}
\end{figure}

In the canonical ensemble with fixed charge, the potential, which is the
free energy of the system presents the thermodynamic behavior of a system in
a standard approach. In order to calculate the free energy of a
gravitational system one may evaluate the Euclidean on-shell action.
Moreover, to make the action well-defined and finite, one should add the
Gibbons-Hawking boundary term and counterterms to the bulk action. Also as
we are working in canonical ensemble we should consider a boundary term for
electromagnetic field namely $I_{s}$ to fix charge on the boundary
\begin{eqnarray}
I_{s} &=&-\frac{p}{4\pi} \int\sqrt{-\gamma} e^{-2p\alpha%
\Phi(r)}(-F)^{p-1}n_{\mu}F^{\mu\nu}A_{\nu}.  \label{Is}
\end{eqnarray}

\begin{eqnarray}
I &=&I_{bulk} +I_{GH}+I_{ct}+I_{s} .  \label{Ic}
\end{eqnarray}
\begin{eqnarray}
G&=&\frac{r_{+}}{4} +\frac{2\pi P(\alpha^{2}-1)(\alpha^{2}+1)}{(3+\alpha^{2})%
} b^{2\gamma}r_{+}^{\frac{3+\alpha^{2}}{\alpha^{2}+1}}  \notag \\
&&+\frac{(\alpha^{2}+1)}{4}\frac{(2p-1)(2p+\alpha^{2}+1) 2^{p} q^{2p}p}{%
(2p+\alpha^{2}+p\alpha^{2}) (\alpha^{2}+3-2p)}b^{\frac{-\gamma(2-2p)}{2p-1}}
r_{+}^{-\frac{\alpha^{2}+3-2p}{(2p-1)(\alpha^{2}+1)}} ,  \label{Gcc}
\end{eqnarray}
where $r_{+}$ is understood as a function of pressure and temperature via
equation of state (\ref{eq of state1}). Moreover since we are considering an
extended phase space, we may associate Gibbs free energy with the $G=M-TS$
\cite{Do1}. In this way the Gibbs free energy can be obtained as
\begin{eqnarray}
G&=&\frac{\omega_{n-1}(\alpha^{2}+1)}{16\pi}\left(\frac{(n-2)k}{%
\alpha^{2}+n-2}b^{\gamma(n-3)} r_{+}^{\frac{\alpha^{2}+n-2}{\alpha^{2}+1}}+%
\frac{16\pi P(\alpha^{2}-1)}{(n-1)(n+\alpha^{2})} b^{\gamma(n-1)}r_{+}^{%
\frac{n+\alpha^{2}}{\alpha^{2}+1}}\right)  \notag \\
&&+\frac{\omega_{n-1}(\alpha^{2}+1)}{16\pi}\frac{(2p-1)(2pn-4p+\alpha^{2}+1)
2^{p} q^{2p}p}{\Pi (\alpha^{2}+n-2p)}b^{\frac{-\gamma(n-1-2p)}{2p-1}}r_{+}^{-%
\frac{\alpha^{2}+n-2p}{(2p-1)(\alpha^{2}+1)}} ,  \label{Gc}
\end{eqnarray}%
which reduces to the (\ref{Gcc}) for $n=3$ and also can reduce to the result
obtained for black holes in EMd gravity as $p=1$ \cite{Kamrani}. The
behavior of the Gibbs free energy is shown in Figs. \ref{Fig4}-\ref{Fig5}.
From these figures we see that there is a swallowtail behavior. It means we
have first order phase transition in the system.

\section{Phase transition in grand-canonical ensemble}

\label{Grand}

In this section we investigate the phase transition in grand-canonical
ensemble by fixing the electric potential at infinity. It is worthwhile to
note that, as one expects for linear Maxwell field $(p=1)$, we cannot see
criticality in the grand canonical ensemble while as we shall see below the
system may encounter a critical behavior in case of the power-Maxwell field
with $p\neq1$.

\subsection{Equation of state}

\begin{figure}[tbp]
\epsfxsize=5cm \centerline{\epsffile{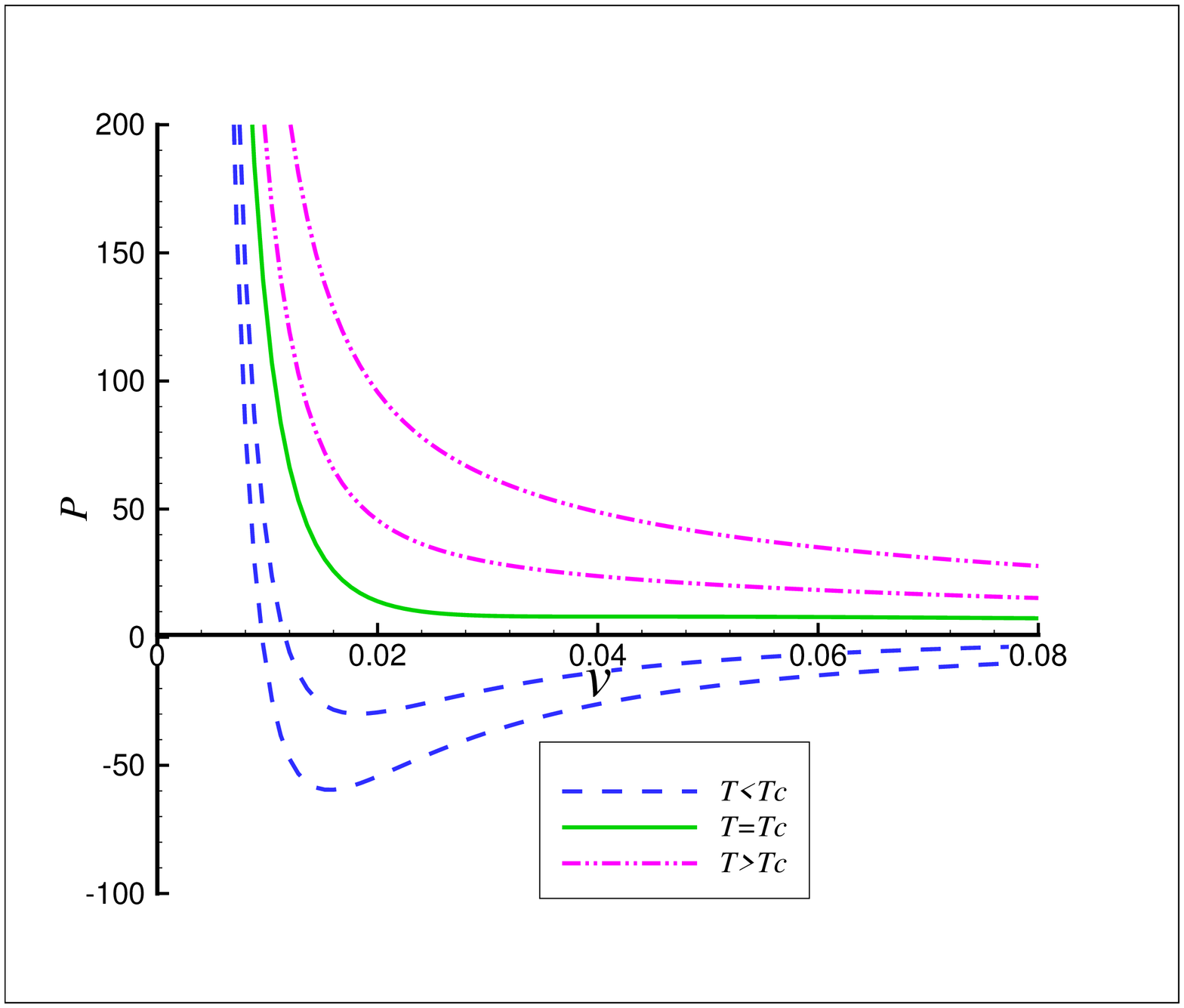}}
\caption{$P-v$ diagram for $b=1$, $n=3$, $U=1$, $k=1$, $p=1.2$ and $\protect%
\alpha =0.3$.}
\label{Fig6}
\end{figure}

\begin{figure}[tbp]
\epsfxsize=5cm \centerline{\epsffile{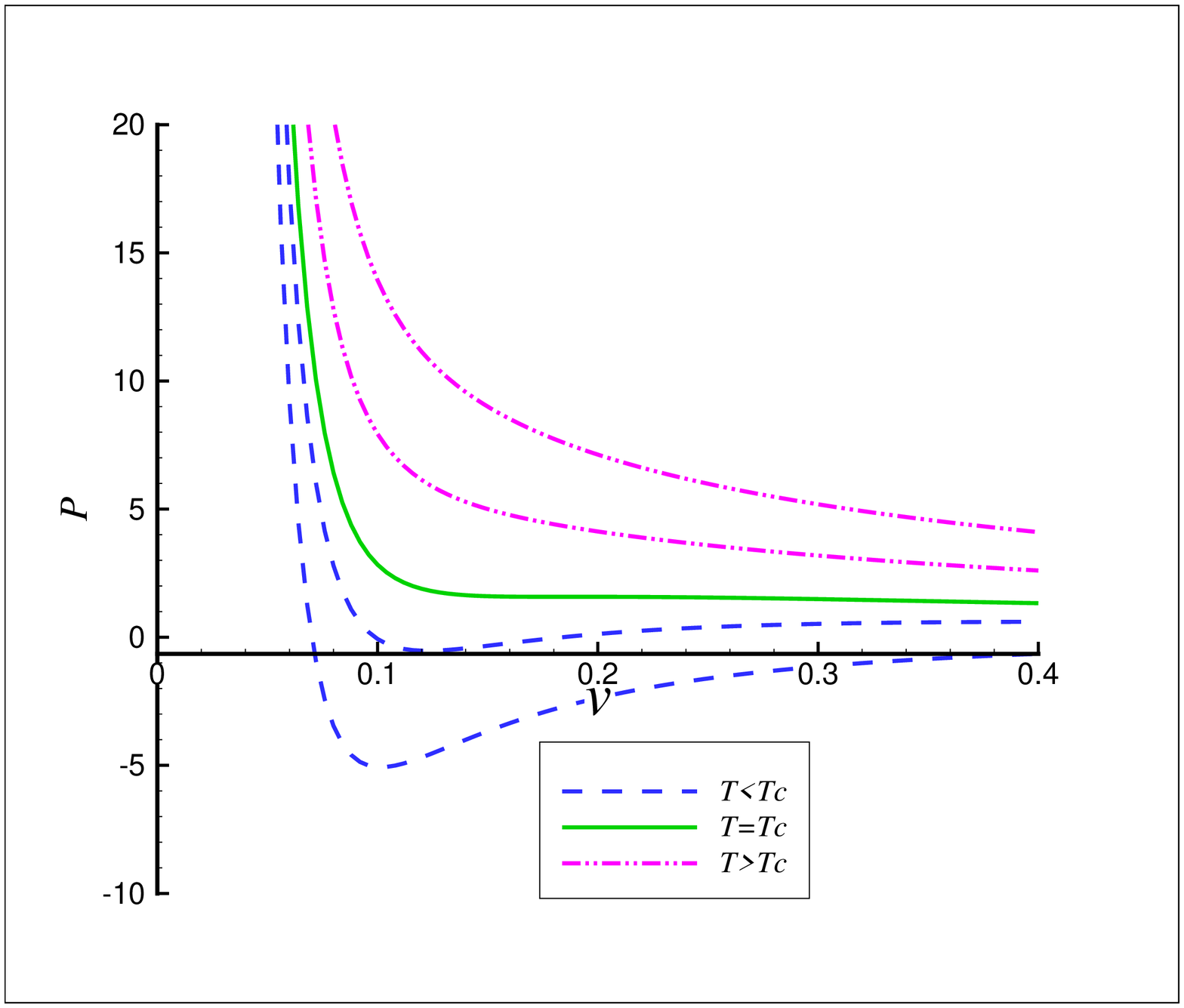}}
\caption{$P-v$ for diagram for $b=1$, $n=3$, $U=1$, $k=1$, $p=2$ and $%
\protect\alpha =0.3$.}
\label{Fig7}
\end{figure}

\begin{figure}[tbp]
\epsfxsize=5cm \centerline{\epsffile{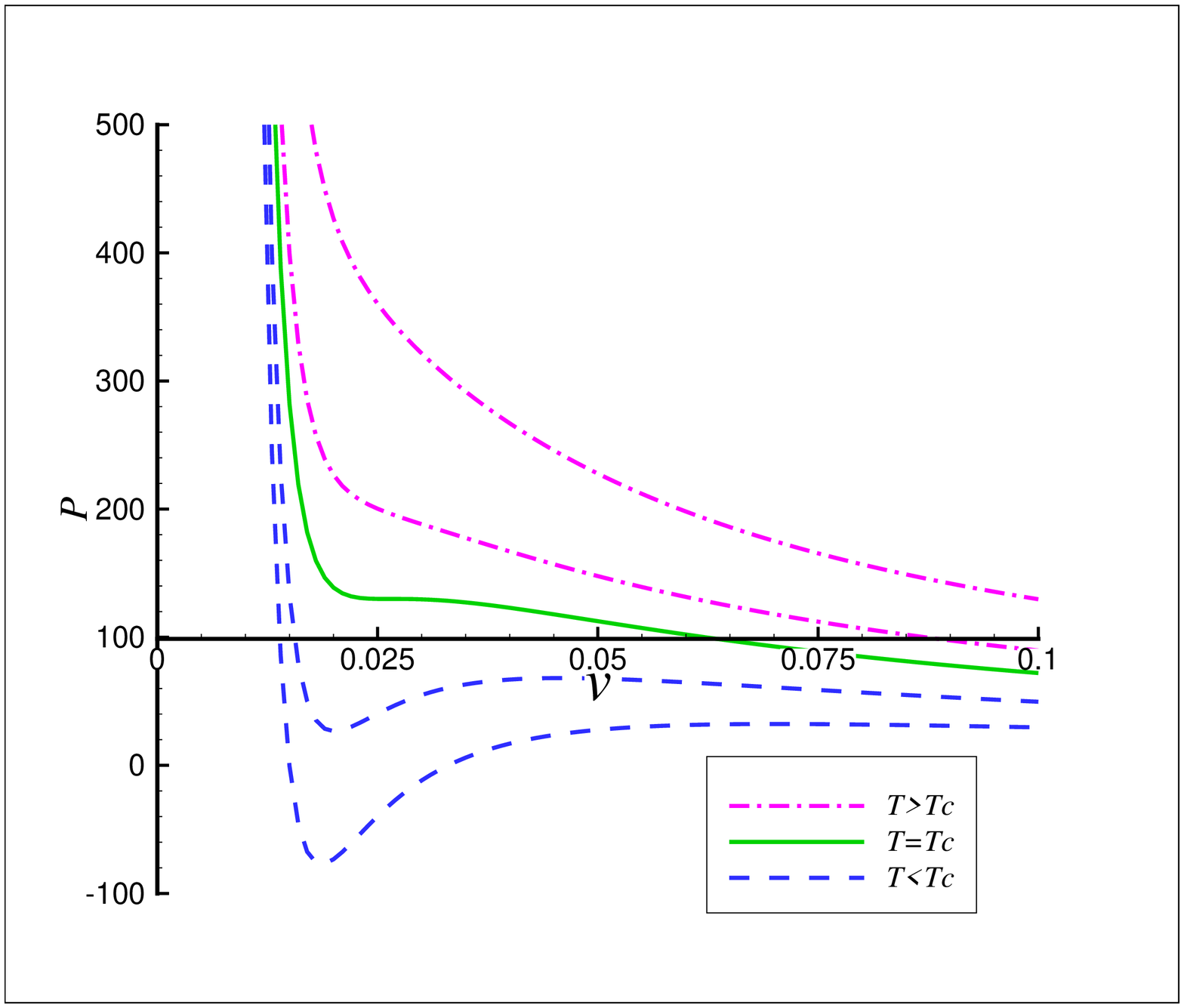}}
\caption{$P-v$ diagram for $b=1$, $n=5$, $U=1$, $k=1$, $p=3$ and $\protect%
\alpha =0.3$.}
\label{Fig8}
\end{figure}

\begin{figure}[tbp]
\epsfxsize=5cm \centerline{\epsffile{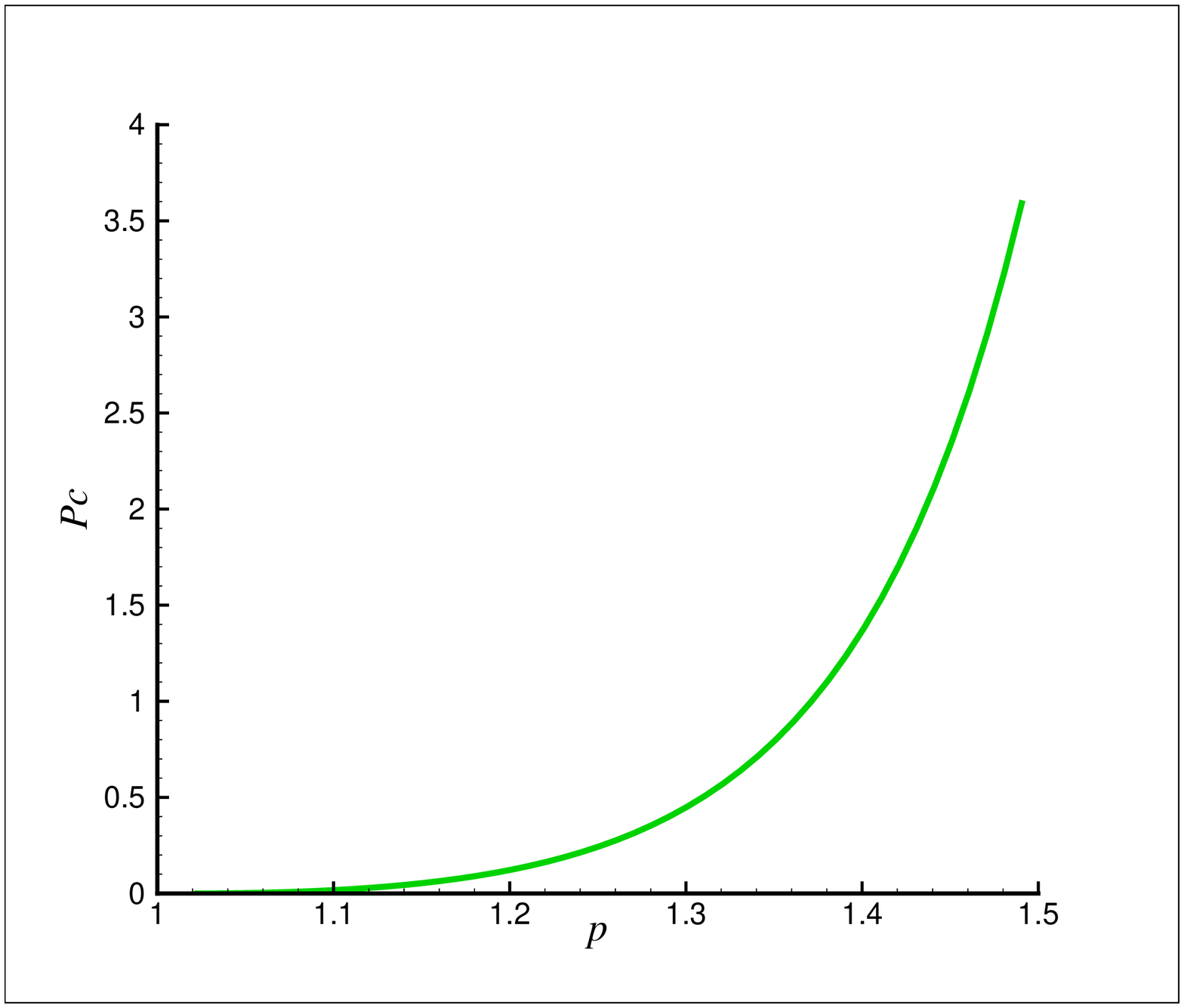}}
\caption{The behavior of $P_{c}$ versus $p$ for $b=1$, $q=1$, $n=4$, $k=1$
and $\protect\alpha =0.3$.}
\label{Fig15}
\end{figure}

To study the critical behavior in grand canonical ensemble, we put
\begin{equation*}
q=\frac{U\Upsilon {r}_{+}^{\Upsilon }}{Cb^{\frac{\left( 2p-n+1\right) \gamma
}{\left( 2p-1\right) }}}.
\end{equation*}%
Using (\ref{eq of state1}) with $r_{+}=v\Gamma $ one may can rewrite
equation of state in the following form
\begin{equation*}
P=\frac{T}{v}-\frac{k(n-2)(1+\alpha ^{2})\Gamma }{4\pi (1-\alpha
^{2})b^{2\gamma }(\Gamma v)^{2-2\gamma }}+\frac{p(2p-1)(1+\alpha ^{2})\Gamma
}{4\pi \Pi }\left( \frac{\sqrt{2}U(n-2p+\alpha ^{2})}{(2p-1)(1+\alpha
^{2})b^{\gamma }c(\Gamma v)^{1-\gamma }}\right) ^{2p}.
\end{equation*}%
In order to compare the critical behavior of the system with van der Waals
gas, we should plot isotherm diagrams. The corresponding $P-v$ diagrams are
displayed in Figs. \ref{Fig6}-\ref{Fig8}.
The critical point can be obtained by solving the following equations
\begin{equation}
\frac{\partial P}{\partial v}\Big|_{T_{c}}=0,\quad \frac{\partial ^{2}P}{%
\partial v^{2}}\Big|_{T_{c}}=0,
\end{equation}%
which leads to
\begin{eqnarray}
v_{c} &=&\frac{1}{\Gamma }Z^{\frac{(\alpha ^{2}+1)}{2(p-1)}}b^{-\alpha ^{2}},
\\
P_{c} &=&\frac{\Gamma k(n-2)(p-1)}{4p\pi }Z^{\frac{-1}{2(p-1)}}, \\
T_{c} &=&\frac{k(n-2)(p-1)b^{-\alpha ^{2}}}{\pi (\alpha ^{2}-1)(\alpha
^{2}+1-2p)}Z^{\frac{\alpha ^{2}-1}{2(p-1)}},
\end{eqnarray}%
where
\begin{equation*}
Z=\frac{2^{p}U^{2p}p^{2}(2p-1)(2p-\alpha ^{2}-1)(n-2p+\alpha ^{2})^{2p}}{%
k(n-2)\Pi \left\{ c(2p-1)(\alpha ^{2}+1)\right\} ^{2p}}.
\end{equation*}%
One can see the behavior of $P_{c}$ and $T_{c}$ in the Figs. \ref{Fig15}-\ref%
{Fig18}.
\begin{figure}[tbp]
\epsfxsize=5cm \centerline{\epsffile{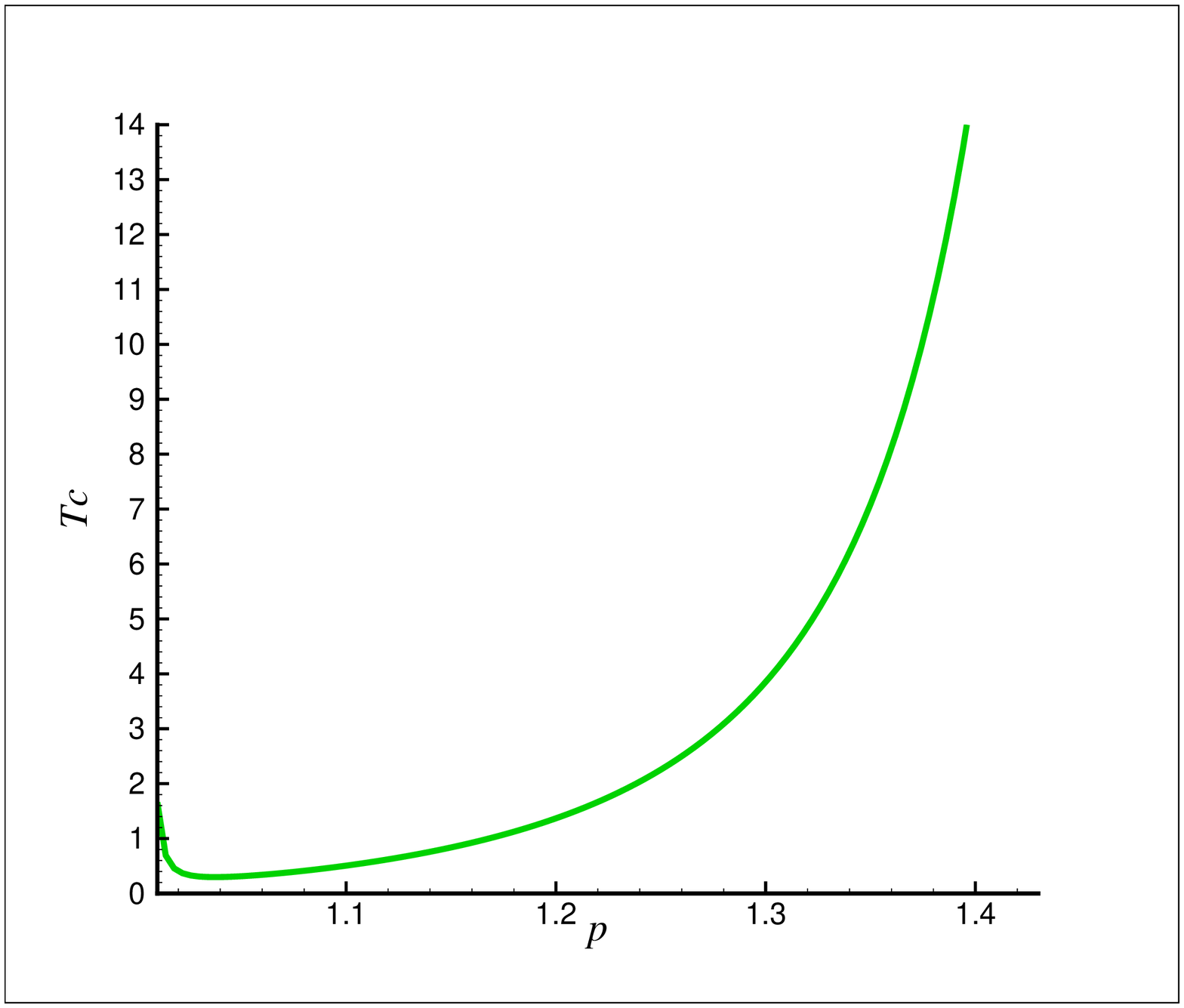}}
\caption{The behavior of $T_{c}$ versus $p$ for $b=1$, $q=1$, $n=3$, $k=1$
and $\protect\alpha =0.3$.}
\label{Fig16}
\end{figure}
\begin{figure}[tbp]
\epsfxsize=8cm \centerline{\epsffile{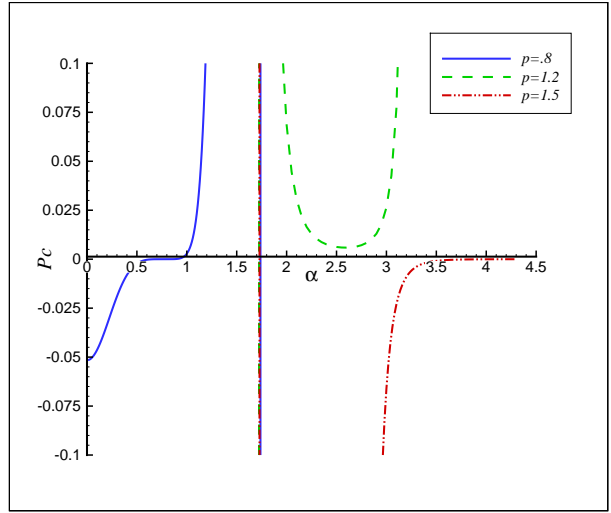}}
\caption{$P_{c}-\protect\alpha $ diagram for $b=1$, $q=1$, $n=3$, $q=1$ and $%
k=1 $.}
\label{Fig17}
\end{figure}
\begin{figure}[tbp]
\epsfxsize=8cm \centerline{\epsffile{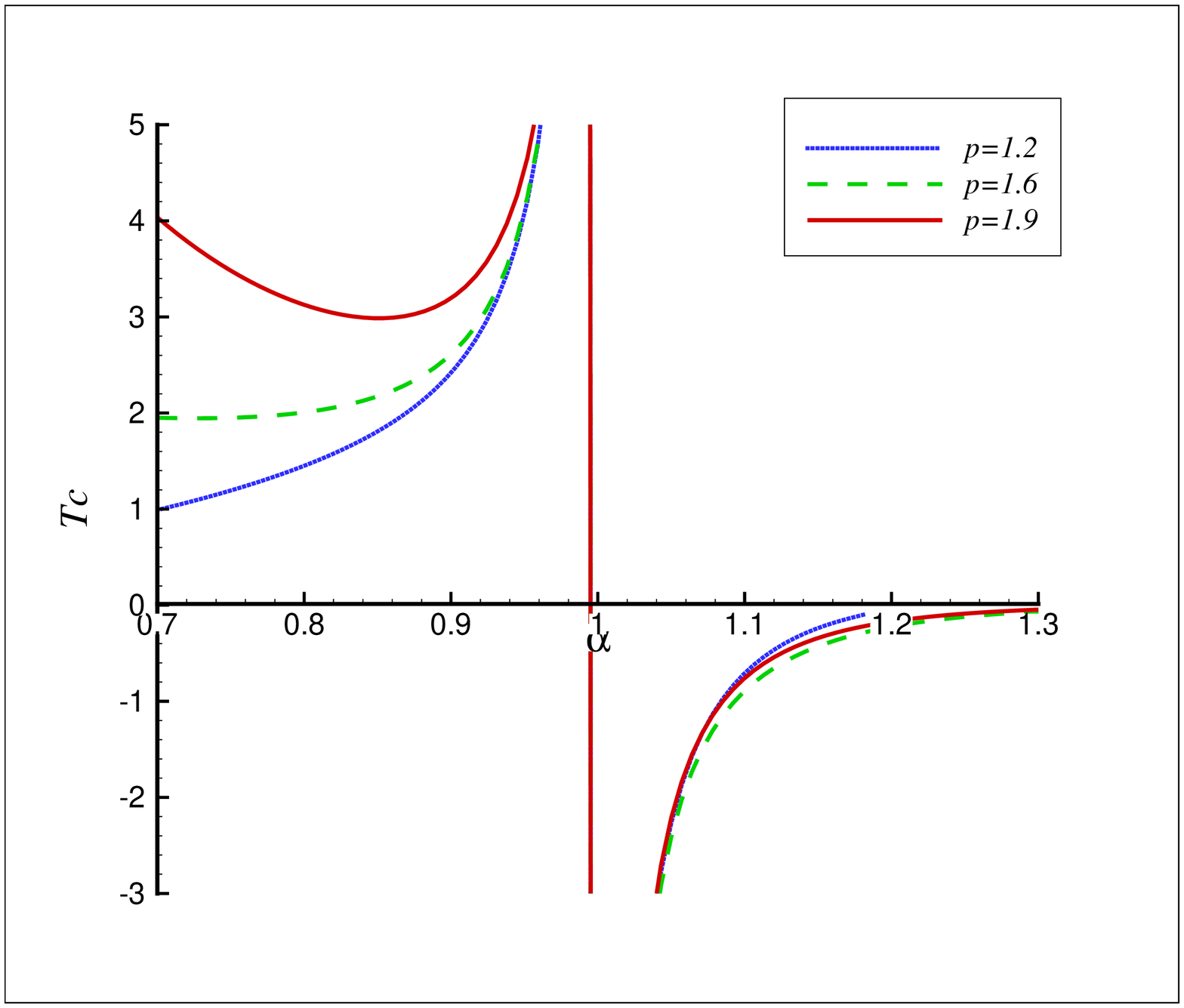}}
\caption{(Color online) $T_{c}-\protect\alpha $ diagram for $b=1$, $q=1$, $%
n=4$, $q=1$ and $k=1 $. }
\label{Fig18}
\end{figure}
Using the above critical values, $\rho _{c}$ is obtained as
\begin{equation}
\rho _{c}=\frac{(\alpha ^{2}+1-2p)(\alpha ^{2}-1)}{4p},
\label{universal ratio2}
\end{equation}%
As one expects, the above $\rho _{c}$ reduces to that of Ref. \cite{Hendi}
as $\alpha $ goes to zero. In the absence of the dilaton field ($\alpha
=0=\gamma $) for $p=2$, it reduces to $3/8$ which is the characteristic of
van der Waals fluid. It is clear that $\rho _{c}$ is independent of $n$.

\subsection{Critical exponents}

Following the method we adopted in the canonical ensemble, we can obtain
\begin{equation}
p=1+At-Bt\omega -C\omega ^{3}+O\left( t\omega ^{2},\omega ^{4}\right) ,
\label{ptw2}
\end{equation}%
with%
\begin{eqnarray}
A &=&\frac{1}{\rho _{c}},\quad B=\frac{1}{\varepsilon \rho _{c}},\quad \\
C &=&\frac{2p}{3(1+\alpha ^{2})^{2}\varepsilon ^{3}}.
\end{eqnarray}
Thus, we obtain the same critical exponent as we found already in the
canonical ensemble
\begin{equation}
\alpha ^{\prime }=0 ,\beta ^{\prime }=1/2 ,\gamma ^{\prime }=1 ,\delta
^{\prime }=3.
\end{equation}

\subsection{Gibbs free energy in grand canonical ensemble}

\begin{figure}[tbp]
\epsfxsize=8cm \centerline{\epsffile{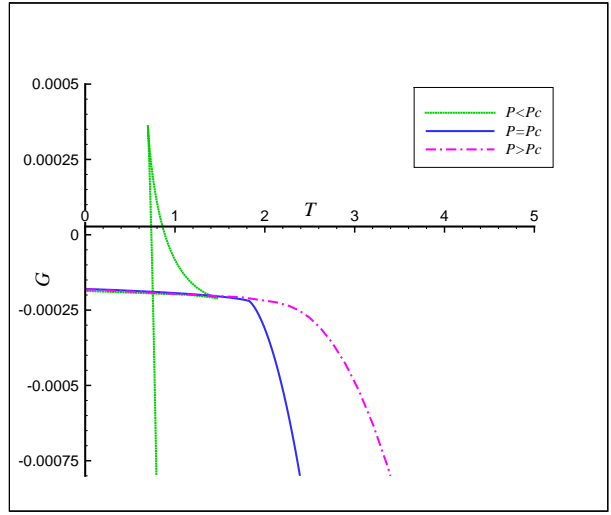}}
\caption{Gibbs free energy versus $T$ for $b=1$, $n=3$, $U=1$, $k=1$, $p
=6/5 $ and $\protect\alpha=0$.}
\label{Fig9}
\end{figure}

\begin{figure}[tbp]
\epsfxsize=8cm \centerline{\epsffile{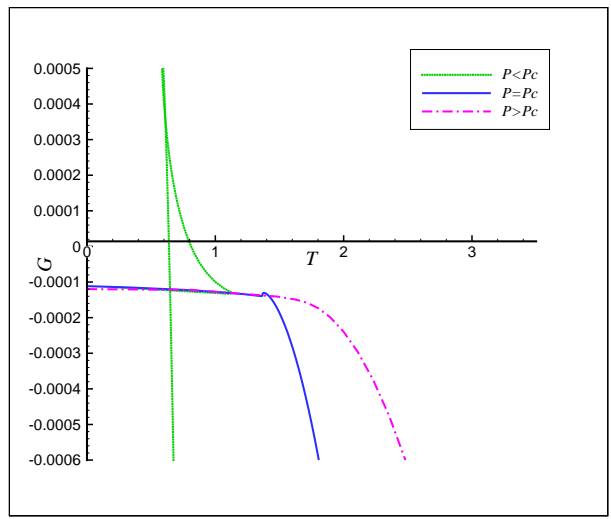}}
\caption{Gibbs free energy versus $T$ for $b=1$, $n=3$, $U=1$, $k=1$, $p
=6/5 $ and $\protect\alpha=0.3$.}
\label{Fig10}
\end{figure}

As we have done for the canonical ensemble, in the grand-canonical ensemble
we can calculate Gibbs free energy by calculating the on-shell action. Since
we are working in the grand canonical ensemble, we should fix electric
potential on the boundary. This implies that we must ignore the surface term
(\ref{Is}). So the total action becomes
\begin{equation}
I =I_{bulk} +I_{GH}+I_{ct} .  \label{Igc}
\end{equation}
\begin{eqnarray}
G&=&\frac{r_{+}}{4} +\frac{2\pi P(\alpha^{2}-1)(\alpha^{2}+1)}{(3+\alpha^{2})%
} b^{2\gamma}r_{+}^{\frac{3+\alpha^{2}}{\alpha^{2}+1}}  \notag \\
&&+ \frac{2^{p}U^{2p}p(3+\alpha^{2}-2p)^{2p-1}(1-\alpha^{2})}{%
4c^{2p}(\alpha^{2}+1)^ {2p-1}(2p-1)^{2p-2}((-1+p)\alpha^{2}-2p)}r_{+}^{\frac{%
3-2p+\alpha^{2}}{(\alpha^{2}+1)}} b^{-\gamma(2p-2)} .  \label{Ggc}
\end{eqnarray}%
It is a matter of calculation to show the above Gibbs energy which is
obtained from action is same as
\begin{equation}
G=M-TS-\frac{\Pi}{2p}QU ,
\end{equation}
in the extended phase space for $n=3$ . The behavior of the Gibbs free
energy is shown in Figs. \ref{Fig9}-\ref{Fig10}.
%%%%%%%%%%%%%%%%%%%%%%%%%%%%%%%%%%%%%%%%%%%%%%%%%%%%%%%%%%%%%%%%%%%%%%%%%%%%%%

\section{Summery and Conclusions}

\label{Sum} In this paper, we have investigated the critical behavior of $%
(n+1)$-dimensional dilaton black holes in the presence of power Maxwell
field. For this purpose, we have calculated the finite action in both
canonical and grand canonical ensembles. We have introduced, for the first
time, the counterterm method for spacetimes with curved boundary in dilaton
gravity. Due to the fact that the asymptotic behavior of the solutions is
not (A)dS, the number of non-vanishing counterterms constructed by the
curvature invariant of the boundary at infinity may be infinite. We found
out that although as $\alpha $ approaches to $\sqrt{n-2}$, the number of
counterterm goes to infinity, the summation of all the divergent terms in
the total action is zero at infinity and therefore one only needs to
calculate the finite terms of the total action (\ref{Itot}). In other words
the number of counterterms which should be calculated in Einstein-dilaton
gravity is the same as that of Einstein gravity and one only needs to
calculate the finite terms of them.

In order to have solutions in Einstein-dilaton gravity in the presence of
power-Maxwell field, one needs three Liouville type potentials \cite{Kord}.
One of the Liouville type potential contains a constant $\Lambda $, which
plays the role of cosmological constant and the others guarantee the
existence of the solution. We extended the phase space by considering the
constant $\Lambda $ to be treated as thermodynamic pressure, and its
conjugate quantity as a thermodynamic volume. By calculating the
thermodynamic quantities, we obtained the Smarr relation, which reduces to
the Smarr relation in the absence of dilaton field given in \cite{Hendi} and
in the limit $p=1$ it reduces to those of \cite{Kamrani}. After constructing
the Smarr relation, we used the pressure and Hawking temperature to build
the equation of state in the canonical and grand canonical ensembles. Then,
we plotted $P$-$v$ isotherm diagrams. These figures show the analogy between
our system and the van der Walls fluid, with the same phase transition in
both ensembles. Interstingly enough, we found that in contrast to the
RN-AdS, EMd and EBId black holes which has the phase transition only in
canonical ensemble, dilaton black holes with power-Maxwell gauge field admit
the critical behaviour in both canonical and grand canonical ensembles. We
also found that the critical behavior can be occurred only for black holes
with spherical horizon ($k=1$). Then, we obtained the critical pressure,
volume and temperature both for the canonical and grand canonical ensembles
and by using them we calculate the action and Gibbs free energy. We have
considered the behavior of the Gibbs free energy and found that there is a
swallowtail behavior for Gibbs free energy as a function of temperature in
both ensembles which shows there is a first order small-large black holes
phase transition in the system. Finally, we calculated the critical
exponents and found that while the critical quantities are different in two
ensembles, these exponents are the same and they are the same as van der
Waals system. This, implies that the inclusion of nonlinear electrodynamics,
dilaton field or extra dimensions do not change the critical exponents.

Finally, we would like to mention that in this work we considered the
power-Maxwell field as the gauge field. It is worth investigating the
effects of dilaton on the critical behavior of black holes in the presence
of other nonlinear gauge field such as logarithmic and exponential
electrodynamics.
%%%%%%%%%%%%%%%%%%%%%%%%%%%%%%%%%%%%%%%%%%%%%%%%%%%%%%%%%%%%%%
\acknowledgments{We thank Shiraz University Research Council. This
work has been supported financially by Research Institute for
Astronomy and Astrophysics of Maragha, Iran.}
%%%%%%%%%%%%%%%%%%%%%%%%%%%%%%%%%%%%%%%%%%%%%%%%%%%%%%%%%%%%%%%%%%%%%%%


\begin{thebibliography}{99}
\bibitem{Do1} B. P. Dolan, Class. Quant. Grav. \textbf{28}, 235017 (2011).
%\textit{Pressure and volume in the first law of black hole thermodynamics}%

\bibitem{Ka} D. Kastor, S. Ray, and J. Traschen, Class. Quant. Grav. \textbf{%
26}, 195011 (2009).
%\textit{Enthalpy and the Mechanics of AdS Black Holes}%

\bibitem{Do2} B. Dolan, Class. Quant. Grav. \textbf{28}, 125020 (2011).
%\textit{The cosmological constant and black hole equation of state},%

\bibitem{Do3} B. P. Dolan, Phys. Rev. D \textbf{84}, 127503 (2011).
%\textit{Compressibility of rotating black holes},%

\bibitem{Ce1} M. Cvetic, G. W. Gibbons, D. Kubiznak, and C. N. Pope, Phys.
Rev. D \textbf{84}, 024037 (2011).
%\textit{%Black hole enthalpy and an entropy inequality for the thermodynamic volume},%

\bibitem{Ur} M. Urana, A. Tomimatsu, and H. Saida, Class. Quant. Grav.
\textbf{26}, 105010 (2009).
%\textit{Mechanical first law of black hole spacetime with cosmological constant and its application to Schawarzchils-de Sitter spacetime}%

\bibitem{MannRN} D. Kubiznak and R. B. Maan, J. High Energy Physics, \textbf{%
07}, 033 (2012). %\textit{P-V criticality of charged Ads black holes }%

\bibitem{MannBI} Sh. Gunasekaran, D. Kubiznak and R. B. Mann, J. High Energy
Physics, \textbf{11}, 110 (2012).
%\textit{%Extended phase space thermodynamics for charged and rotating black holes and Born-Infeld vacuum polarization}%

\bibitem{Sherkat} M. B. Jahani Poshteh, B. Mirza and Z. Sherkatghanad, Phys.
Rev. D 88, 024005 (2013).
%\textit{Phase transition, critical behavior, and critical exponents of Myers-Perry black holes}%

\bibitem{Sherkat1} Z. Sherkatghanad, B. Mirza, Z. Mirzaeyan and S. A.
Hosseini Mansoori, arXiv:1412.5028.
%\textit{Critical behaviors and phase transitions of black holes in higher order gravities and extended phase spaces}%

\bibitem{Rabin} R. Banerjee and D. R Roychowdhury, Phys. Rev. D \textbf{85},
044040 (2012);\newline
R. Banerjee, D. Roychowdhury, Phys. Rev. D \textbf{85}, 104043 (2012).
%\textit{Critical behavior of Born Infeld AdS black holes in higher dimensions}%

\bibitem{Zou} De. Ch. Zou, Sh.-J. Zhang and B. Wang, Phys. Rev. D \textbf{89}%
, 044002 (2014).
%\textit{\ Critical behavior of Born-Infeld AdS black holes in the extended phasespace thermodynamics}%

\bibitem{Hendi} S. H. Hendi, and M. H. Vahidinia, Phys. Rev. D \textbf{88},
084045 (2013);\newline
S. H. Hendi, S. Panahiyan and B. Eslam Panah, arXiv:1410.0352.
% \textit{Extended phase space thermodynamics and P-V criticality of black holes with nonlinear source%}%

\bibitem{John} C. V. Johnson, Class. Quant. Grav. \textbf{31}, 225005 (2014)
C. O. Lee, Phys. Let. B \textbf{09} (2014) 046.
%\textit{The extended thermodynamic phase structure of Taub-Nut and Taub-Bolt}% %\textit{The extended thermodynamic properties of Taub-NUT/bolt-AdS spaces}%

\bibitem{De} De. Zou, Y. Lio and B. Wang, Phys. Rev. D. textbf{90}, 044063
(2014).
%\textit{critical behavior of charged Gauss-Bonnet ADS black holes in the grand canonocal ensamble}%

\bibitem{Xi} J. X. Mo and W. B. Liu, Eur. Phys. J. C. \textbf{74}, 2836
(2014).
% \textit{P-V Criticality of Topological Black Holes in Lovelock-Born-Infeld Gravity}%

\bibitem{Ren} R. Zhao, H. H. Zhao, M. S. Ma and L. C. Zhang, Eur. Phys. J. C
\textbf{73}, 2645 (2013).
%textit{On thecritical phenomena and thermodynamics of charged topologicaldilaton AdS black holes}%

\bibitem{Kamrani} M. H. Dehghani, S. Kamrani and A. Sheykhi, Phys. Rev. D.
\textbf{90} 104020,(2014).
%\textit{P-Vcriticality of charged dilatonic black holes}%

\bibitem{dayyani} M. H. Dehghani, A. Sheykhi and Z. Dayyani, Phys. Rev. D
\textbf{93}, 024022 (2016).
%\textit{%Critical behavior of Born-Infeld dilaton black holes}%

\bibitem{kord50} M. Hassaine and C. Martinez, Phys. Rev. D \textbf{75},
027502 (2007).

\bibitem{kord51} H.A. Gonzalez, M. Hassaine and C. Martinez, Phys. Rev. D
\textbf{80}, 104008 (2009).

\bibitem{kord52} S. H. Hendi, Eur. Phys. J. C \textbf{69}, 281 (2010).

\bibitem{kord53} S. H. Hendi, Class. Quantum Gravity \textbf{26}, 225014
(2009).

\bibitem{kord54} S. H. Hendi, Phys. Lett. B \textbf{677} (2009) 123.

\bibitem{kord55} H. Maeda, M. Hassaine and C. Martinez, Phys. Rev. D \textbf{%
79}, 044012 (2009).

\bibitem{kord56} A. Sheykhi, Phys. Rev. D \textbf{86}, 024013 (2012).

\bibitem{kord57} A. Sheykhi and S. H. Hendi, Phys. Rev. D \textbf{87},
084015 (2013).

\bibitem{kord58} M.H. Dehghani, A. Sheykhi and S. E. Sadati, Phys. Rev. D
\textbf{91}, 124073 (2015).

\bibitem{kord59} M. Kord Zangeneh, A. Sheykhi and M. H. Dehghani, Phys. Rev.
D \textbf{92}, 024050 (2015).

\bibitem{kord60} S. H. Hendi and H. R. Rastegar-Sedehi, Gen. Relat. Gravit.
\textbf{41}, 1355 (2009).

\bibitem{Kord} M. Kordzangeneh, A. Sheykhi and M. H. Dehghani, Phys. Rev. D
\textbf{91}, 044035 (2015).

\bibitem{BY} J. D. Brown and J. W. York, Phys. Rev. D \textbf{47}, 1407
(1993).

\bibitem{23mehdizade} J. D. Brown, J. Creighton and R .B. Mann, Phys. Rev. D
\textbf{50}, 6394 (1994).

\bibitem{24mehdizade} I. S. Booth, R .B. Mann, Phys. Rev. D \textbf{59},
064021 (1999).

\bibitem{25mehdizade} V. Balasubramanian and P. Kraus, Commun. Math. Phys.
\textbf{208}, 413 (1999).

\bibitem{26mehdizade} P. Kraus, F. Larsen and R. Siebelink, Nucl. Phys. B
\textbf{563}, 259 (1999).

\bibitem{27mehdizade} V. Balasubramanian and P. Kraus, Commun. Math. Phys.
\textbf{208}, 413 (1999).

\bibitem{28mehdizade} M. Hennigson and K. Skenderis, J. High Energy Phys.
\textbf{7}, 023 (1998).

\bibitem{30mehdizade} S. Nojiri and S. D. Odintsov, Phys. Lett. B \textbf{444%
}, 92 (1998).

\bibitem{31mehdizade} S. Nojiri, S. D. Odintsov and S. Ogushi, Phys. Rev. D
\textbf{62}, 124002 (2000).

\bibitem{32mehdizade} M. H. Dehghani, Phys. Rev. D \textbf{66}, 044006
(2002).

\bibitem{33mehdizade} M. H. Dehghani, Phys. Rev. D \textbf{65}, 124002
(2002).

\bibitem{34mehdizade} M. H. Dehghani and A. Khodam-Mohammadi, Phys. Rev. D
\textbf{67}, 084006 (2003).

\bibitem{Dilaton} M. H. Dehghani, Phys. Rev. D \textbf{71}, 064010 (2005);%
\newline
M. H. Dehghaniand N. Farhangkhah, Phys. Rev. D \textbf{71}, 044008 (2005);%
\newline
A. Sheykhi, M. H. Dehghani, N. Riazi and J. Pakravan, Phys. Rev. D \textbf{74%
}, 084016 (2006);\newline
M. H. Dehghani, A. Sheykhi and S. H. Hendi, Phys. Lett. B \textbf{659}, 476
(2008).

\bibitem{CHM} K. C. K. Chan, J. H. Horne and R. B. Mann, Nucl. Phys. \textbf{%
B447}, 441 (1995).

\bibitem{Shey} A. Sheykhi, Phys. Rev. D \textbf{76}, 124025 (2007).

\bibitem{Cai3} R. G. Cai and K. S. Soh, Phys. Rev. D \textbf{59}, 044013
(1999).
\end{thebibliography}
\end{document}